\documentclass[10pt]{article}

\newtheorem{tm}{Theorem}

\newtheorem{lw}{Law}

\begin{document}
\pagestyle{plain}
\begin{center}
\vskip10mm
   {\Large {\bf Structure behind Mechanics I: Foundation}}
\vskip7mm
\renewcommand{\thefootnote}{\dag}
{\large Toshihiko Ono}\footnote{ 
e-mail: BYQ02423@nifty.ne.jp \ \ or \ \ 
tono@swift.phys.s.u-tokyo.ac.jp}
\vskip5mm
\par\noindent
{\it 703 Shuwa Daiich Hachioji Residence,\\
4-2-7 Myojin-cho, Hachioji-shi, Tokyo 192-0046, Japan}
\vskip5mm
\end{center}
\vskip10mm
\setcounter{footnote}{0}
\renewcommand{\thefootnote}{\arabic{footnote}}

\begin{abstract}
This paper 
proposes a basic theory on
physical reality
 and
a new foundation for quantum mechanics
and classical mechanics.
It presents a scenario  
not only to solve the problem
of the arbitrariness on the
operator ordering for the quantization procedure,
but also to clarify how the classical-limit occurs.
This paper is the first of the three papers
into which the previous paper quant-ph/9906130
has been separated for readability.\\
\begin{center}
{\it Submitted to Found. Phys.}
\end{center}
\end{abstract}
\vskip10mm

\section{INTRODUCTION}

Seventeenth century saw
 Newtonian mechanics,
published as "{\it Principia:
Mathematical principles of natural philosophy},"
the first attempt
to understand this world under few principles
rested on observation and experiment.
It bases itself  on the concept of the {\it force}
acting on a body and on the laws relating it with the motion.
In eighteenth century,
Lagrange's {\it analytical
mechanics},
originated by Mautertuis' theological work,
 built the theory of motion
on an analytic basis, and replaced forces by potentials;
in the next century, Hamilton 
completed the foundation of analytical mechanics
on the principle of least action in stead of Newton's laws.
Besides,
Maxwell's theory of the electromagnetism
has the Lorentz invariance
inconsistent with the invariance
under Galilean transformation,
that Newtonian mechanics obeys.
Twentieth century dawned with
Einstein's relativity 
changing the ordinary belief on the nature of time,
to reveal the four-dimensional {\it spacetime} 
structure of the world.
Relativity
improved Newtonian mechanics
based on the fact that the speed of light $c$
is an invariant constant,
and revised the self-consistency
of the classical mechanics.
Notwithstanding such a revolution,
 Hamiltonian mechanics was
still effective not only for Newtonian mechanics
but also for the Maxwell-Einstein theory,
and the concept of
{\it energy and momentum}
played the most important role in the physics
instead of force for Newtonian mechanics.

Experiments, however, indicated
that microscopic systems seemed not to obey
such classical mechanics so far. Almost one century has passed
since Planck found his constant $h$;
and almost three fourth since
Heisenberg \cite{Heisenberg},  Schr\"odinger
\cite{Schrodinger} and their contemporaries
constructed
the basic formalism of quantum mechanics
after the early days of Einstein and Bohr.
The quantum mechanics
based itself on the concept of {\it wave functions} 
instead of classical energy and momentum,
or that of operators called as observables.
This mechanics
reconstructed the classical field theories
except the general relativity.
Nobody  denies how quantum mechanics,
especially quantum electrodynamics,
succeeded in twentieth century
and developed in the form of the standard model for the
quantum field theories
through the process to find new particles in the nature.

Quantum mechanics,
however,
seems to have left some fundamental open problems
on its formalism and its interpretation:
the problem
on the ambiguity of the operator ordering
in quantum mechanics \cite{Groenwald,van Hove},
which is crucial to quantize the Einstein gravity for instance,
and that on the reality,
which seems incompatible with the causality \cite{EPR,Bell,Aspect}.
These difficulties come from the problem
how and why quantum mechanics relates itself with classical mechanics:
the relationship between the quantization that
constructs quantum mechanics based on
classical mechanics
and the classical-limit
that induces classical mechanics
from quantum mechanics as an approximation
with Planck's constant $h$ taken to be zero;
the incompatibility between
the ontological feature
of classical mechanics and 
the epistemological feature
of quantum mechanics
in the Copenhagen interpretation \cite{Bohr}.

Now,
this paper
proposes a basic theory on 
physical reality, and
introduces a foundation
for quantum mechanics and classical mechanics,
named as {\it protomechanics}, 
that is motivated in the previous letter \cite{Ono}.\footnote{ The author 
of paper \cite{Ono}, "Tosch Ono," 
is the same person as that of the present paper,
"Toshihiko Ono."}
It also
attempts to revise the
nonconstructive idea
that the basic theory of motion is valid in a
way independent of the describing 
scale, though the quantum mechanics has once
destroyed 
such an idea that Newtonian mechanics held in eighteenth century.
The present theory
supposes that a field or a particle $X$
on the four-dimensional spacetime
has its internal-time $\tilde o_A(X)$ relative
to an domain $A$ of the spacetime,
whose boundary and interior
represent the present and the past, respectively.
It further considers that
object $X$ also has the 
external-time $\tilde o_A^*(X)$ relative
to  $A$  which is the internal-time of all the rest but
$X$ in the universe.
Object $X$ gains the actual existence on  $A$
if 
and only if the internal-time
coincides with the external-time:
\begin{equation}\label{AX=XA}
\tilde o _A(X)= \tilde o_A^*(X) .
\end{equation}
This condition discretizes or quantizes the ordinary time passing from 
the past to the future,
and
enables the deterministic structure of
the basic theory to produce the
nondeterministic characteristics of
quantum mechanics.
The both sides of relation (\ref{AX=XA}) further
obey the variational principle as
\begin{equation}
 \label{dAX=0/dXA=0}
\delta \tilde o_A\left( X \right)  = 0
\ \ \ , \ \ \ \
\delta \tilde o^*_{A } (X) = 0 .
\end{equation}
This relation reveals a geometric
structure behind Hamiltonian mechanics
based on the modified  Einstein-de Broglie relation,
and produces the conservation law
of the emergence-frequency of a particle or a field
based on the introduced quantization law of time.
The obtained mechanics,
protomechanics,
rests on the concept of the 
{\it synchronicity}\footnote{This naming
of synchronicity is originated by Jung \cite{Jung&Pauli}.}
instead of energy-momentum or wave-functions,
that synchronizes two intrinsic local clocks
located at different points in the space of
the objects on a present
surface in the spacetime.
It will finally
solve the problem
on the ambiguity of the operator ordering,
and also give a self-consistent interpretation
of quantum mechanics as an ontological theory.

The next section
explains the basic laws on reality
as discussed above,
and leads to the protomechanics in
Section 3, that produces
the conservation laws of momentum
and that of emergence-frequency.
Section 4 presents the dynamical
construction 
for the introduced protomechanics
by utilizing the group-theoretic method
called Lie-Poisson mechanics (consult {\it APPENDIX}).
It provides the difference between classical
mechanics
and quantum mechanics as that of their {\bf function spaces}:
the function space of the observables 
for quantum
 mechanics includes
that for  classical mechanics;
 the dual space of the emergence-measures
for classical 
mechanics includes
that
for quantum
mechanics, viceversa.
A brief statement of
the conclusion immediately follows.

The present paper shall
leave to the following paper \cite{SbMII}
the detail proof how the protomechanics
deduces classical mechanics and quantum mechanics,
since such proof needs a intricate mathematical technique
strayed from the present context;
and it will demonstrate there still valid for the description
of a half-integer spin against
the ordinary belief
that the existence of such spin 
averts realistic approaches to the quantum mechanics
from the completeness.
It also has to leave to another paper \cite{SbMIII}
the concluded implication
how the present theory 
gives
a self-consistent interpretation
for quantum mechanics,
since such discussion needs a philosophical
background  beyond  the scope of the present paper;
and it will  further prove there to provide the 
semantics of the regularization
 in a quantum field theory,
the quantization of a phenomenological system,
the causality in quantum mechanics and
 the origin of the thermodynamic irreversibility
under the new insight.\footnote{
The paper of quant-ph/9906130
contains the information not only in the present paper
but also in the following two papers \cite{SbMII,SbMIII}.}

The following diagram
illustrates the construction of the present paper.

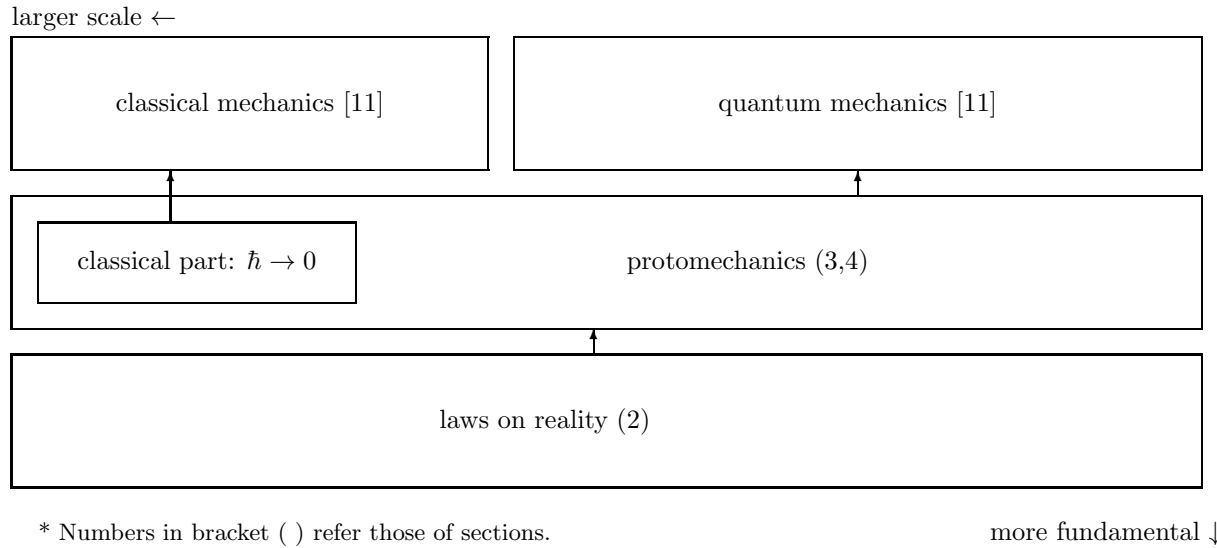
\begin{picture}(480,220)(15,20)

\put(10,160){\framebox(180,50){classical mechanics \cite{SbMII}}}

\put(200,160){\framebox(260,50){quantum mechanics \cite{SbMII}}}

\put(10,100){\framebox(450,50){\ \ \ \ \ \ \ \ \ \ \ \ \ \ \ \ \ \  
\ \ \ \ \ \ \ \ \ \ \ \ \ \  protomechanics (3,4)}}
\put(20,110){\framebox(120,30){classical part: $\hbar \to 0$}}

\put(10,40){\framebox(450,50){laws on reality (2) \ \ \ \ \ \ \ \ \ \ \ \ \ }}

\put(230,90){\vector(0,1){10}}
\put(70,140){\vector(0,1){20}}
\put(330,150){\vector(0,1){10}}

\put(10,215){larger scale $ \leftarrow $}
\put(380,20){more fundamental $ \downarrow $}
\put(20,20){{\small * Numbers in bracket $( \ )$ refer those of sections.}}
\end{picture}
{}\\
{}\\

In this paper, 
I will
use Einstein's rule
in the tensor calculus
for Roman indices'
$i, j, k \in {\bf N}^N$
and Greek indices' 
$\nu , \mu  \in {\bf N}^N$,
and not for Greek indices' 
$\alpha , \beta , \gamma \in {\bf N}^N$.
Consult
the brief review 
 on Lie-Poisson mechanics
in {\it APPENDIX}.
In addition,
notice that the basic theory
uses so-called {\bf c-numbers},
while it will also utilize {\bf q-numbers} 
to deduce the quantum mechanics in 
\cite{SbMII} for the help of calculations.\footnote{
Such distinction between c-numbers and q-numbers
 does not play an important
role in the present theory.}

\section{LAWS ON REALITY}

Let $M^{(4)}$  represent 
the spacetime, being a four-dimensional oriented
$C^{\infty }$ manifold,
that has the topology
or the family $\tilde {\cal O}={\cal O}_{M^{(4)}}  $ of its open subsets, 
the topological $\sigma $-algebra
${\cal B}\left( {\cal O}_{M^{(4)}} \right) $,
and the volume measure $v^{(4)}$ induced from the metric $g$
on $M^{(4)}$.\footnote{Spacetime $M^{(4)}$ 
may be endowed with some additional structure.}
We shall certainly choose an
arbitrary domain $A\in \tilde {\cal O} $
in the discussion below,
but we are interested in
the case that domain
$A$ represents the past at a moment whose boundary
$\partial A$ is a three-dimensional present hypersurface
in $M^{(4)}$.

The space ${\tilde M} $ represents
that of the objects
whose motion will be described,
and has a projection operator $ \chi _A:{\tilde M} \to {\tilde M}$
for every domain $A\in \tilde {\cal O}$
such that $\chi_A^2=  \chi _A$.
Every object $X\in {\tilde M} $ has
its own domain $D(X)$ such that
\begin{equation}
\chi_{D(X) \setminus A } (X) =X \ \ \ \ \Longleftrightarrow 
\ \ \ \ D(X) \cap A =\emptyset .
\end{equation}
In particle theories,
${\tilde M} $ is identified with the space
 of all the one-dimensional
timelike mani-folds or curves in $M^{(4)}$,
where 
 $\chi_A \left( l\right) =l \cap A $ for every domain $A$
and $D(l) =l$.
In field theories,
the space $\Psi \left( M^{(4)}, V\right) $
of the complex valued or
${\bf Z}_2$-graded fields over $M^{(4)}$
such that $\psi^{(4)} \in \Psi \left( M^{(4)}, V\right)  $ is
a mapping $\psi^{(4)}  : M^{(4)} \to V$
for a complex valued or
${\bf Z}_2$-graded vector space $V$.
Mapping $\chi_A $ satisfies that
 $\chi_A  \left( \psi^{(4)} \right) (x) 
=  \psi^{(4)}  (x) $ if $x\in A$
and that
$\chi_A  \left( \psi^{(4)} \right) (x) 
=  
0 $ if $x\not\in A$, 
and $D(\psi^{(4)})$
gives  the support of $\psi^{(4)}$:
$D(\psi^{(4)}) = supp (\psi^{(4)})$.

In addition,
let us consider
the set ${\cal D} (\tilde M)$ of 
all the differentiable mapping from $\tilde M$ to itself
and the set ${\cal D} ( M^{(4)})$ of 
all the diffeomorphisms of spacetime $M^{(4)}$.
In particle theories,
set ${\cal D}(\tilde M )$ will be regarded as set ${\cal D}(M^{(4)})$;
and,
in field theories, it is the set 
of all the linear transformations of
a field
such that
$\Phi \left( \psi^{(4)}\right) = \psi^{(4)}+ \phi^{(4)} $.

Now,
let us assume
that an object
has its own internal-time 
relative to a domain of the spacetime.

\begin{lw}\label{Variation A}
For every domain $A\in \tilde {\cal O}$,
the mapping
 $\tilde o_A : 
 {\tilde M}  \to S^1$  has an action $S _A :  \tilde M  \to {\bf R} $
and equips  an object $X\in {\tilde M}$
with the internal-time
$\tilde o_A(X) $:
\begin{equation}\label{original Feynamm's rule}
\tilde o _A \left(  X \right) = e^{i 
S_A ( X ) }  .
\end{equation}
\end{lw}
For particle theories,
a one-dimensional submanifold or a curve
$l\subset M^{(4)}$
 represents
the nonrelativistic motion for a particle 
 such that
$\left( t, x(t) \right)  \in l 
 $ for $t\in T$,
 where
$M^{(4)}$ is the Newtonian spacetime $M^{(4)}= T\times 
M^{(3)}$ for the
Newtonian time $T\subset {\bf R}$
and the three-dimensional Euclidean space $M^{(3)}$;
thereby, it 
has
the following action 
for the ordinary Lagrangian $L :TM\to {\bf R}$:
\begin{equation}\label{nonrelativistic action}
S _A\left( l 
\right)  = {\bar h}^{-1} \int_ {l\cap A} d t  \ L \left( x(t)
, {{dx(t)}\over {dt}}
\right)  ,
\end{equation}
where 
$\bar h= h/ {4\pi } $ or
$=\hbar / 2$ for Planck's constant $h$ ($\hbar =h/ 2\pi $).
The
relativistic motion
of a free particle whose mass is $m$
has the following action for the proper-time $\tau \in {\bf R}$:
\begin{equation}\label{relativistic action}
S _A\left( l 
\right)  = {\bar h}^{-1} \int_ {l\cap A} d \tau  \ mc^2  .
\end{equation}
 For field theories,
field variable $X =\psi^{(4)} $ over spacetime $M^{(4)}$
 has the 
following action
for the Lagrangian density ${\cal L}_M$ of matters:
\begin{equation}\label{standard action}
S _{A}\left( \psi^{(4)}
\right)   ={1\over {\bar h c}} \int_{A}
  d v^{(4)}   \left( y  \right) \ {\cal L}_M  \left(  \psi^{(4)}  (y),
d\psi^{(4)}  (y)
\right)   ,
\end{equation}
where $ v^{(4)} $ is the volume measure of $M^{(4)}$.
In the standard field theory,
 $ \psi ^{(4)}$ is a set of ${\bf Z}_2$-graded fields
over spacetime $M^{(4)}$,
the Dirac field for fermions,
 the Yang-Mills field for gauge bosons
and other field under consideration.
For  the
Einstein gravity,
the
Hilbert action
includes
the metric tensor $g$ on $M^{(4)}$ 
with
a cosmological constant $\Lambda \in {\bf R}$:
\begin{eqnarray}
\label{Hilbert action}\nonumber
S _{A}\left(  \psi^{(4)}, g
\right)   &=&{1\over {\bar h c}} \int_{A}
  d v_g^{(4)}   \left( y  \right) \ {\cal L}_M\left(  
\psi  ^{(4)}(y),
d\psi  ^{(4)}(y)
\right) \\
& \ & \ \ \ -  {1\over {\bar h c}} \int_{A}
  d v_g^{(4)} \left( {{c^4} \over {16 \pi G}} R_g +\Lambda \right)
 - {2\over {\bar h c}} \int_{\partial A}
  d v_g^{(3)} {{c^4} \over {16 \pi G}} K_g  ,
\end{eqnarray}
where $R_g$ and $K_g$ are the
four-dimensional and the extrinsic three-dimensional  scalar curvatures on 
domain $A$ and on its boundary ${\partial A}$;
and $G$ is the Newton's constant of gravity.
The last
term of (\ref{Hilbert action}) is necessary to
produce the correct Einstein equation for gravity \cite{Wald}.

Let us now consider
the subset ${\cal D}_A(\tilde M)$ of set ${\cal D}(\tilde M)$
such that
every element $ \Phi\in {\cal D}_A(\tilde M)$
satisfies
$ \chi_{D(X)\setminus A}(\Phi (X) ) = X $,
and assume it as a infinite-dimensional Lie group.
In particle theories,
set ${\cal D}_A(\tilde M)$ is the set ${\cal D}_A(M)$
of all the diffeomorphisms
of $M$ such that
$\Phi (l) \setminus A = l \setminus A $;
and,
in filed theories,, it is the set 
of all the linear transformations of
a field
such that
$\Phi \left( \psi^{(4)}\right) = \psi^{(4)}+ \phi^{(4)} $ 
for an element $\phi^{(4)}\in \Psi \left( M^{(4)}, V\right) $
and that $ \phi^{(4)}(x)=0$  if $x\not\in A $.
Mapping  $\tilde o_A $ may have the symmetry
under a transformation $\Phi\in
{\cal D} (\tilde M)$  such that
it satisfies 
the following relation for every pair $(A,X)$:
\begin{equation}\label{sym}
\tilde o_{ A}\left( \Phi (X)\right) = \tilde o_A (  X) .
\end{equation}
Such symmetry 
verifies the existence of the conserved charge.

Object $X$ and all the rest but $X$
composes
the universe $U$.
The internal-time $\Pi_A (U)$ of 
universe $U$ relative to domain $A$
would be separated into two parts:
\begin{equation}
\Pi_A (U) = \tilde o_A (X) \cdot \tilde o^*_A ( X) .
\end{equation}
Let us call $\tilde o^*_A ( X)\in S^1$ as the external-time
of $X$ relative to $A$.
Thus, the external-time of universe $U$
would always be unity: $\Pi^*_A (U) =1$.

\begin{lw}\label{Variation A2}
For every domain $A\in \tilde {\cal O}$,
the mapping
 $\tilde o_A^* : 
 {\tilde M}  \to S^1$  has an action $S^* _A :  \tilde M  \to {\bf R} $
and equips  an object $X\in {\tilde M}$
with the external-time
$\tilde o^*_A(X) $:
\begin{equation}\label{original Feynamm's rule}
\tilde o ^* _A\left(  X \right) = e^{i 
S_A^*  ( X ) }  .
\end{equation}
\end{lw}
Let us also introduce the mapping $\tilde s_A  \left(  \tilde o  \right)
:
\tilde M  \to S^1$
that relates mappings $\tilde o^*_A $ and $\tilde o_A$:
\begin{equation}\label{original Emergencer}
\tilde o^*_{A} (X) 
= \tilde o _A\left( X \right)
\cdot \tilde s_A  \left(   \tilde o  \right) (X) .
\end{equation}
It has a function $R_A\left( \tilde o \right) $ such that
\begin{equation}
\tilde s_A\left(   \tilde o \right) (X)
= e^{i R_A\left(  \tilde o \right) (X) } .
\end{equation}
There is also the mapping $ \tilde s_{A }^*  \left(  \tilde o ^* \right) :
\tilde {\cal O}   \to S^1$:
\begin{equation}
\tilde o^*_{A} (X)
\cdot \tilde s_{A }^* \left( \tilde o^* \right) (X)
 = \tilde o _A\left( X \right) .
\end{equation}
Mapping  $\tilde \eta ^*_A$ may have the symmetry
under a transformation  $\Phi\in
{\cal D} (\tilde M)$ 
 such that
it satisfies 
the following relation for every pair $(A,X)$:
\begin{equation}
\tilde o^*_{A}\left( \Phi (X)\right) = \tilde o^*_A (  X) .
\end{equation}
If mapping  $\tilde \eta_A$ also
has symmetry (\ref{sym}) for the same transformation $\Phi $,
they must satisfy the following invariance:
\begin{equation}
\tilde s_A \left( \tilde o\right) \left( \Phi (X)\right) = 
\tilde s_A \left( \tilde o\right)  (  X) 
\ \ \ , \ \ \ \ 
\tilde s^*_A \left(
\tilde o^*\right)\left( \Phi (X)\right) = 
\tilde s^*_A \left(\tilde o^*\right) (  X) .
\end{equation}

The following law further
supplies
the condition that an
object 
has the actual existence on a domain of the spacetime.

\begin{lw}\label{internal-external}
Object $X\in \tilde M$ 
has actual existence on domain $A\in \tilde {\cal O} $
when the internal-time coincides with the external-time:
\begin{equation}\label{compa}
\tilde o^*_{A} (X) = \tilde o _A\left( X \right) .
\end{equation}
\end{lw}
Relation (\ref{compa})
requires the following quantization
condition:
\begin{equation}\label{uncer}
\tilde s_A \left(  \tilde o \right) (X) =1,
\end{equation}
or equivalently,
\begin{equation}
\tilde s_{A}^* \left(  \tilde o^* \right) (X)=1,
\end{equation}
which 
quantizes spacetime $M^{(4)}$
for 
an object $X\in \tilde M$.

For the space $ d_A(\tilde M)$ of all the infinitesimal generators
of ${\cal D}_A(\tilde M)$,
let us consider
an arbitrary element
$\Phi_{\epsilon }\in {\cal D}_A(\tilde M)$,
differentiable by parameter $\epsilon \in {\bf R}$:
\begin{equation}
\lim_{\epsilon \to 0} {{d\Phi_{\epsilon }}\over {d\epsilon }}
 \circ \Phi_{\epsilon }^{-1} 
= \xi \in d_A(X) .
\end{equation}
Thus, we can
introduce
the variation $\delta $ 
as follows:
\begin{eqnarray}
\left\langle 
i \tilde o_A \left(  X \right)  ^{-1 } 
\delta \tilde o_A \left(  X \right) , \xi\right\rangle
&=& i \tilde o_A \left(  X \right) ^{-1 } 
\left. {d\over {d\epsilon }}\right\vert_{\epsilon =0}
\tilde o_A \left(  \Phi _{\epsilon }(X) \right) ,\\
\left\langle 
i \tilde o_A^* \left(  X \right)  ^{-1 } 
\delta \tilde o_A^* \left(  X \right) , \xi\right\rangle
&=& i \tilde o_A^* \left(  X \right) ^{-1 } 
\left. {d\over {d\epsilon }}\right\vert_{\epsilon =0}
\tilde o_A^* \left(  \Phi _{\epsilon }(X) \right) 
\end{eqnarray}
where $\langle \ \ , \ \ \rangle :
 d^*_A(\tilde M) \times  d_A(\tilde M) \to {\bf R}$
is the natural pairing for the dual space $ d^*_A(\tilde M)$
of $ d_A(\tilde M)$.
This variation satisfies the variational principle
of the following law.

\begin{lw}\label{variational}
Object $X\in \tilde M$ must
 satisfy the variational principle
for every domain $A\in \tilde {\cal O} $:
\begin{equation}
 \label{v-pr}
\delta \tilde o _A (X)  = 0
\ \ \ , \ \ \ \
\delta \tilde o^*_{A } (X) = 0 .
\end{equation}
\end{lw}
Thus, Law 4  keeps
Law 3 under the above variation,
and also has the following
expression:
\begin{equation}
 \label{v-pr2}
\delta \tilde s _A\left( \tilde o \right) (X)  = 0
\ \ \ , \ \ \ \
\delta \tilde s ^*_A\left( \tilde o^* \right) (X) = 0 .
\end{equation}

Now, we will
consider 
the mapping
${\cal P} : T \to \tilde {\cal O} $
for 
the time $T\subset {\bf R}$ of an observer's clock $T$.
Domain
${\cal P}(t)$ and
its boundary $\partial  {\cal P} (t )
= \overline{{\cal P} (t )}\setminus {\cal P} (t )$
represent
 the {\it past} and the {\it present}
at time $t \in T$,
where $\overline{A}$ is the closure of $A\in \tilde {\cal O}$;
and it satisfies the following conditions:
\begin{enumerate}
\item
for every $X\in \tilde M$,
$
 t_1 < t_2  \in T \ \ \Rightarrow \ \
{\cal P} (t_1 )  \cap D (X )  \subset {\cal P} (t_2 )  \cap D (X ) $
 (ordering);
\item 
for every $X\in \tilde M$,
the present $
\partial  {\cal P} (t ) \cap D (X ) 
$ is a spacelike hypersurface in $M^{(4)}$
for every time $t\in T$ (causality).
\end{enumerate}
From Law 3, object $X$  emerges into the world  at
time $t \in T$
when it satisfies
\begin{equation}\label{unit}
 \tilde s_{{\cal P}(t)} \left( \tilde o \right) (X)=1\ .
\end{equation}
This condition
of the emergence
determines when object $X$ interacts with all the 
rest in the world,
and discretizes time $T$
in Whitehead's philosophy 
\cite{Whitehead}. In other words,
what a particle or a field $ X$ 
gains actual existence or
emerges into the world, here,
means that it becomes exposed to or has the possibility to interact 
with the other elements or with the ambient world
excluded from the description.
Such occasional influences from the unknown factors
can break the deterministic feature of 
the above description; and it would cause the irreversibility
in general as considered in elsewhere \cite{SbMIII}.
The emergence further allows
the observation
of a particle or a field
through an experiment even if the device or its environment is
included in the description \cite{SbMIII}.
Besides, the variational principle of Law 4
produces the equation of motion and
the conservation
of the frequency of such emergence in the next section.

\section{Foundation of Protomechanics}

Let us consider
the development of present 
$\partial {\cal P}(t)$ for short time $T=(t_i,t_f) \subset {\bf R}$,
keeping the following description
without the appearance of singularity;
and suppose 
that the time interval extends long enough
to keep the continuity of time
beyond the discretization in the previous section,
where such discretization
would only affect the property of the emergence-measure,
defined below, corresponding to the density matrices
in quantum mechanics.
Assume that present
$\partial {\cal P}(t)$ is diffeomorphic to
a three dimensional manifold $M^{(3)}$
by a diffeomorphism
$\sigma_t :M^{(3)} \to \partial {\cal P}(t)  $
for every $t\in T$.
It induces a corresponding mapping
$\tilde \sigma_t : \tilde M \to M $
for the space $M$
that is three-dimensional
physical space $M^{(3)}$ for particle theories
or the space $M= \Psi (M^{(3)},V)$ 
of all the $C^{\infty }$-fields 
over $M^{(3)}$ for field theories.
For particle theories, mapping
$ \tilde \sigma_t  $ is defined as
$\tilde \sigma_t (l) =
\sigma_t ^{-1} \left( l\cap  \sigma_t (M^{(3)}) \right) $
for a curve $l\subset M^{(4)}$;
for field theories, it is defined as
$ \tilde \sigma_t (\psi^{(4)})  =
\psi^{(4)} \circ \sigma_t $ for a
field $\psi^{(4)} $.

Let us assume that space $M$ is a $C^{\infty }$ manifold
endowed with an appropriate
topology and the induced topological $\sigma $-algebra.\footnote{
$M $ is assumed as an ILH-manifold modeled by the Hilbert space
endowed with an inverse-limit topology (consult \cite{Omori}).}
We will denote the tangent space as $TM$
and the cotangent space $T^*M$;
and we shall consider
the space of all the vector fields over $M$ as $X(M)$
and that 
of all the 1-forms over $M$ as $\Lambda^1(M)$.
To
add a one-dimensional cyclic freedom $S^1$ at
each point of $M$ introduces
the 
$S^1$-fiber bundle $E (  M ) $ over $M$.\footnote{The 
introduced freedom may
{\it not} directlly represent what is corresponding to the
{\it local clock} in Weyl's sense
or the {\it fifth-dimension}
in Kaluza's sense \cite{Kaluza} for the 
four-dimensional spacetime  $M^{(4)}$, if related.}
Fiber $S^1$ represents an 
intrinsic clock of a particle or a field,
which is located at every point on $M$.
For
the space $\Gamma  [ E (  M ) ] $
of all the  global sections
of $E (  M ) $,
every element $ 
\eta \in \Gamma  [ E ( M)   ] $
now represents the system
that a particle or a field
belongs to and carries with,
and a synchronization of every two clocks
located at different points in space $ M$.

For past ${\cal P}(t)$ such that  
$\partial {\cal P}(t) = \sigma_t (M^{(3)})$,
there is an mapping $o_t : TM \to {\bf R}$ such that
every initial position $(x_0, \dot x_0 )\in TM$ 
has an object $X\in \tilde M_{{\cal P}}$
satisfying the following relation
for $x_t=\tilde \sigma_t (X)$:
\begin{equation}\label{def eta}
o_t \left( x_t,  \dot x_t \right) = 
\tilde o_{{\cal P}(t)} \left( X \right)  .
\end{equation}
For the velocity field $v_t \in X(M)$ 
such that $v_t \left( x_t\right) = 
 {{dx_t}\over {dt}} $, we will introduce a
section $\eta_t \in \Gamma \left[ E(M)\right] $
and call it {\it synchronicity} over $M$:
\begin{equation}\label{def eta}
\eta_t (x)=o_t \left( x, v_t (x) \right)   .
\end{equation}
The Lagrangian $L_t^{TM}: TM \to {\bf R}$ 
 characterizes
the speed of the internal-time:
\begin{equation}\label{internal velocity}
L^{TM}_t\left( x_t, {{dx_t}\over {dt}} \right) 
 = -i\bar h   o_t \left( x_t,
{{dx_t}\over {dt}} \right)^{-1} {d\over {dt}}o_t \left( x_t,
{{dx_t}\over {dt}} \right) .
\end{equation}
Since relation (\ref{internal velocity})
is valid for every initial conditions of position $
(x_t,\dot x_t) \in TM$,
it determines the time-development of synchronicity $\eta_t $
in
the following way
for the Lie derivative ${\cal L}_{v_t}$
by velocity field
$v_t \in X(M)$:
\begin{equation}\label{internal Lie velocity}
L_t^{TM}\left( x, v_t (x)\right) =
-i {\bar h}  \eta_t   (x) ^{-1}\left( {\partial \over {\partial t }}
+ {\cal L}_{v_t} \right) \eta_t   (x)  .
\end{equation}

Let us now consider
the mapping $ p :
\Gamma [ E (  M )   ]
\to \Lambda^1(M)
$ satisfying  the following relation:
\begin{equation}
p \left( \eta_t \right) =
- i \bar h  \eta_t ^{ -1}
d \eta_t .
\end{equation}
If 
the energy
$ E_t \left( \eta_t \right) :TM \to {\bf R}$
is defined as 
\begin{equation}
E_t \left( \eta_t \right)\left( x \right) =  i\bar h
\eta_t   (x)   ^{-1 }
{{ \partial   }\over {\partial t }} 
\eta_t   (x) ,
\end{equation}
condition (\ref{internal Lie velocity})
satisfies the following relation:
\begin{equation}
\label{H-L rel.}
E_t \left( \eta_t \right)
 (x)  
=
 v_t  (x) \cdot p \left( \eta_t \right) (x)   - 
L^{TM}_t \left( x, v_t(x)  \right) .
\end{equation}

Attention to
the following calculation by definition (\ref{internal Lie velocity}):
\begin{equation}\label{calc-1}
-i\bar h {\partial \over {\partial v }}
\left\{ o_t \left( x, v_t( x)   \right)^{-1}
\left( {\partial \over {\partial t }}
+ {\cal L}_{v_t} \right) o_t \left( x, v_t( x)  \right) \right\} =
{{\partial L^{TM}_t}\over {\partial v}}\left( x,
v_t  (x)
\right) .
\end{equation}
Since variational principle
(\ref{v-pr}) in
Law 1 implies that $ o_t \left( x, \dot x  \right) $
is invariant under the variation
of $\dot x $ at every point $(x, \dot x)$, i.e.,
\begin{equation}\label{variation>>motion}
{\partial \over {\partial \dot x }}o_t \left( x, \dot x  \right)  = 0
\ \ \ \ \Longleftrightarrow  \ \ \ \ 
{\partial \over {\partial v }}o_t \left( x, v_t (x) \right)    = 0
\end{equation}
then formula (\ref{calc-1})
has the following different expression:
\begin{eqnarray}
\nonumber
-i\bar h {\partial \over {\partial v }} \left\{
o_t \left( x, v_t( x)   \right)^{-1}
\left( 
{{\partial } \over {\partial t }}
+ {\cal L}_{v_t} 
\right) 
o_t \left( x, v_t( x) \right) \right\}
 &=&
{{\partial } \over {\partial v }} \left\{
v_t(x)\cdot  p \left( \eta_t   \right) (x) \right\}
\\ \label{calc-2}
&=&
p \left( \eta_t   \right) (x ) .
\end{eqnarray}
Equations (\ref{calc-1}) and (\ref{calc-2})
leads to the {\it modified}
Einstein-de Broglie relation, that was
$p=h / \lambda $ for Planck's constant $h = 2\pi \hbar $
and wave number $\lambda $
in quantum mechanics:
\begin{equation}\label{p-lagrange}
 p \left( \eta_t   \right) (x) = 
{{\partial L^{TM}_t}\over {\partial v}}\left( x,
v_t  (x)
\right) .
\end{equation}
Notice that this
relation (\ref{p-lagrange})
produces the Euler-Lagrange equation
resulting from
the classical least action principle:
\begin{eqnarray}
\label{E-L}
d L^{TM}_t  \left( x,
v_t ( x)
\right)
- \left( {{\partial }\over {\partial t}}+{\cal L}_{v_t} \right)
{{\partial L^{TM}_t}\over {\partial v}}\left( x,
v_t ( x)
\right) &=&0 \\
\label{E-L4}
\Longleftrightarrow \ \ \ \ \ 
{ { \partial L^{TM}_t  }\over {\partial x^j}}\left( x_t,
\dot x_t
\right)
- {{d}\over {dt}}
{{\partial L^{TM}_t}\over {\partial \dot x^j}}\left( x_t,
\dot x_t
\right) &=&0 \ ;
\end{eqnarray}
thereby, relation (\ref{p-lagrange})
is stronger condition than the classical relation  (\ref{E-L4}).

Under the modified Einstein-de Brogie relation (\ref{p-lagrange}),
relation (\ref{H-L rel.})
gives the Legendre transformation and
introduces Hamiltonian $H_t^{T^*M}$ 
as a real function on cotangent space $T^*M$
such that 
\begin{equation}
E_t \left( \eta_t   \right)
 (x)   =
H_t^{T^*M}\left( x ,p \left( \eta_t   \right) (x)  \right) .
\end{equation}
This satisfies the first equation of Hamilton's canonical
equations of motion:
\begin{equation}\label{velocity-H}
v_t (x)= {{\partial H^{T^*M}_t}\over {\partial p}}\left( x,
p\left( \eta_t  \right) (x)
\right) .
 \end{equation}
Solvability
$\left[ {\partial \over {\partial t}}, d \right] =0$ further
leads to
 the second equation
of Hamilton's canonical equations
of motion:
\begin{equation}\label{solvability}
{\partial \over {\partial t}} p \left( \eta_t   \right) (x)  = 
- d H^{T^*M}_t   \left( x,  p \left(  \eta_t   \right) (x) \right)  ,
\end{equation}
which is equivalent to
equation (\ref{E-L}) of motion under condition (\ref{p-lagrange}).
If Lagrangian $L_t^{TM} $ satisfies
\begin{equation}\label{cons-T}
{{\partial L_t^{TM} }\over {\partial t}}=0 \ ,
\end{equation}
then
equations (\ref{velocity-H}) and (\ref{solvability})
of motion 
prove the conservation
of energy:
\begin{equation}\label{conservation of energy}
\left( {{\partial }\over {\partial t}}+{\cal L}_{v_t}\right) H_t^{T^*M}
\left( x,  p \left(  \eta_t   \right) (x) \right)  =0 .
\end{equation}

On the other hand,
the mapping $ \tilde s_{{\cal P}(t)}\left( \tilde o\right) $
induces a
mapping
$s_t ( o_t   ) :TM \to S^1 $
such that every initial position $(x_0,\dot x_0) \in TM$
has an object $X\in \tilde M_{{\cal P}}$
satisfying the following relation:
\begin{equation}
s_t (  o_t )  \left( x_t, {{dx_t}\over {dt}}\right) = \tilde s_{{\cal P}(t)}
\left(  \tilde o\right) (X).
\end{equation}
For velocity field $v_t$,
we can define the following section
$\varsigma_t \left( \eta_t   \right)  \in \Gamma \left[ E(M)\right] $
and call it {\it shadow} over $M$:
\begin{equation}
\varsigma_t \left( \eta_t   \right) (x) =
s_t (  o_t )  \left( x , v_t(x)\right) .
\end{equation}
Condition (\ref{unit}) of emergence now
has the following form:
\begin{equation}
s_t\left(  o_t\right)  \left( x_t, {{dx_t}\over {dt}}\right) =1
\ \ \ \ \Longleftrightarrow \ \ \ \
\varsigma_t \left( \eta_t   \right) (x) =1,
\end{equation}
when synchronicity
$\eta_t $ comes across the 
section $\eta_t^*=\eta_t
\cdot \varsigma_t\left( \eta_t\right)  $ at position $ x \in M$.
Let us introduce 
the function $T_t( o_t )^{TM}: TM \to {\bf R}$ such that
\begin{equation}\label{internal velocity 2}
T_t(o_t )^{TM}\left( x_t, {{dx_t}\over {dt}} \right)  
= -i  \bar h
 s_t(o_t )\left( x_t, {{dx_t}\over {dt}} \right) ^{-1}
{d\over {dt}}s_t(o_t ) \left( x_t, {{dx_t}\over {dt}} \right) 
.
\end{equation}
Since relation (\ref{internal velocity 2})
is valid for every initial conditions of position $x_t \in M$,
it determines the time-development of shadow $\varsigma_t\left( \eta_t\right)  $
in
the following way
for the Lie derivative ${\cal L}_{v_t}$
by the velocity field
$v_t \in X(M)$ such that $v_t \left( x_t\right) =  {{dx_t}\over {dt}} $:
\begin{equation}\label{s-Lie}
T_t(o_t )^{TM}\left( x, v_t(x) \right)  =
 -i  \bar h \varsigma_t \left( \eta_t   \right) ^{-1}
\left\{ 
{{\partial }\over {\partial t}}
+ {\cal L}_{v_t}\right\} \varsigma_t \left( \eta_t   \right) \ .
\end{equation}
In stead of Hamiltonian
for a synchronicity,
we will consider
the emergence-frequency $f_t \left( \eta_t \right)
:  M \to {\bf R}$ for a shadow such that
\begin{equation}
 2 \pi \bar h    f_t
\left( \eta_t \right) (x)  = i 
\bar h \varsigma_t \left( \eta_t   \right) (x)  ^{-1} {{
\partial  }\over {\partial t}} \varsigma_t \left( \eta_t   \right) (x) ,
\end{equation}
which represents
the frequency that a particle or a field
emerges into the world.
Condition (\ref{s-Lie})
satisfies the following relation:
\begin{equation}
\label{H-L rel.2}
  2 \pi \bar h     f_t \left( \eta_t \right)
(x)=  
v_t  (x) \cdot p \left( s_t \left( \eta_t \right) \right) (x)    
- T_t (o_t )^{TM}\left( x, v_t(x)  \right) .
\end{equation}

Variational principle
(\ref{v-pr2}) from
Law 4 implies that $ s_t(o_t ) \left( x, \dot x  \right) $
is invariant under the variation
of $\dot x $ at every point $(x, \dot x)$, i.e.,
\begin{equation}\label{variation>>emergence}
{\partial \over {\partial \dot x }}s_t (o_t ) \left( x, \dot x  \right)  = 0
\ \ \ \ \Longleftrightarrow \ \ \ \
{\partial \over {\partial v }}
s_t (o_t ) \left( x, v_t( x ) \right)   = 0,
\end{equation}
which
leads to the following relation
corresponding to the {\it modified}
Einstein-de Broglie relation for synchronicity $\eta_t$:
\begin{equation}\label{p-lagrange: emergence}
 p \left( \varsigma _t\left( \eta_t \right)  \right) (x) = 
{{\partial T^{TM}_t (o_t)}\over {\partial v}}\left( x,
v_t  (x)
\right) .
\end{equation}
Relation (\ref{p-lagrange: emergence})
proves the conservation
of emergence-frequency
in the same way as 
relation (\ref{p-lagrange}) proved that of energy
(\ref{conservation of energy}):
\begin{equation}
\left( {{\partial }\over {\partial t}}+{\cal L}_{v_t}\right) f_t
\left( \eta_t\right) (x)  =0 .
\end{equation}
Notice that emergence-frequency $ f_t\left( \eta_t \right)
 $ can
be negative as well as positive,
and that it 
produces a similar property of
the Wigner function
for a
wave function in quantum mechanics \cite{SbMII}.

In addition,
the probability measure $\tilde \nu $
on $\tilde M $
induces
 the probability measure $\nu_t $ 
on $M$ at time $t\in T$
such that
\begin{equation}
 d\nu_t \left( x_t,
{{dx_t}\over {d t}}
\right)   =  d\tilde \nu (X),
\end{equation}
that represents the ignorance of the initial
position in $M$;
thereby it satisfies the  conservation law:
\begin{equation}
{{d}\over {d t}}d\nu_t  \left( x_t,
{{dx_t}\over {d t}}
\right)   =0 .
\end{equation}
This relation can be described by using the Lie derivative 
${\cal L}_{v_t}$  as
\begin{equation}
\left( {{\partial }\over {\partial t}}
+{\cal L}_{v_t}\right) d\nu_t \left( x, v_t(x) \right)  =0 .
\end{equation}
Since
the velocity field $v_t$
has relation (\ref{velocity-H}) with
synchronicity $\eta_t$,
we can define the {\it emergence-measure}
 $\mu_t \left( \eta_t \right) $
as the product of  the probability measure
with the emergence-frequency:
\begin{equation}
d\mu_t \left( \eta_t \right) (x)= d\nu_t \left( x, v_t(x) \right)  \cdot 
f_t
\left( \eta_t \right)  (x)  .
\end{equation}
Thus, we will obtain the following equation of motion
for emergence-measure 
$d\mu_t(\eta_t )$:
\begin{equation}\label{quotient}
\left( {{\partial }\over {\partial t}}
+{\cal L}_{v_t}\right) d\mu_t \left( \eta_t \right) = 0.
\end{equation}

Let me summarize
the obtained mechanics or protomechanics
based on equations (\ref{internal Lie velocity})
and (\ref{quotient}) of motion 
with relation (\ref{velocity-H})
 in the following theorem
that this section proved.

\begin{tm}(Protomechanics)
Hamiltonian $H^{T^*M}_t : T^*M \to {\bf R}$
defines the velocity field $v_t \in {\cal X}(M)$
and Lagrangian $L^{TM}_t: TM\to  {\bf R}$ as follows:
\begin{eqnarray}
v_t (x)
&=& {{\partial H^{T^*M}_t}\over {\partial p}}\left( x,
p\left( \eta_t \right) (x)
\right) \\
L^{TM}_t \left( x,
v(x)  \right)
&=&  v  (x) \cdot p \left(\eta _t \right) (x)   -  
H^{T^*M}_t\left( x, p\left(\eta _t \right) (x)  \right) ,
\end{eqnarray}
where mapping
$p: \Gamma [E(M)] \to \Lambda^1 (M)$ satisfies
the modified Einstein-de Broglie relation:
\begin{equation}
p \left(\eta _t \right) =
- i \bar h   \eta_t  ^{ -1}
d \eta_t .
\end{equation}
The equation of motion
is the set of the following equations:
\begin{eqnarray}
\left( {{\partial }\over {\partial t}}
+{\cal L}_{v_t} \right) \eta_t (x)&=&  -i {\bar h}^{-1}
L_t^{TM}\left( x, v_t (x)\right)   \eta_t (x) ,\\
\left( {{\partial }\over {\partial t}}
+{\cal L}_{v_t} \right) d\mu_t \left( \eta_t \right) &=& 0.
\end{eqnarray}
\end{tm}

\section{DYNAMICAL CONSTRUCTION OF PROTOMECHANICS}

Let us 
express the introduced protomechanics 
in the statistical way
for the ensemble of all the synchronicities on $M$,
and
construct
the dynamical description
for the 
collective motion
of the sections of $E(M)$.
Such statistical description realizes
the description within a
long-time interval 
through the introduced relabeling process
so as to change the labeling time,
that is the time for the
initial condition
before analytical problems occur.
In addition,
it clarifies the relationship between
classical mechanics
and quantum mechanics
under the assumption
that the present theory safely
induces them,
and that will be proved in the following paper \cite{SbMII}.\footnote{
In another way,
consult quant-ph/9906130.}
For mathematical simplicity,
the discussion below
suppose that
$M$ is a $N-$dimensional manifold for a finite 
natural number
$N\in {\bf N}$.

The derivative operator $D= \hbar dx^{j}\partial_{j} : T_0^m (M) \to T_0^{m+1} (M)$
($m\in {\bf N}$)
for the space $T^n_0 (M)$ of all the $(0,n)$-tensors on $M$
can be described as
\begin{equation}\label{D-der}
 D^n p(x) = \hbar ^n \left( \prod_{k=1}^{n}\partial_{j_k} p_j(x) \right)
dx^j \otimes \left( \otimes_{k =1}^{n} dx^{j_k} \right) .
\end{equation}
By utilizing this derivative operator $D$,
the following Banach norm
endows the space $\Gamma \left[ E(M) \right] $ of 
all the $C^{\infty }$ sections of $E(M)$
 with a norm topology 
for the family ${\cal O}_{\Gamma \left( E(M) \right) }$
of the induced open balls:
\begin{equation}\label{q-norm}
\left\|   p( \eta  ) \right\| = \sup_{M} 
\sum_{\kappa \in {\bf Z}_{\geq 0} } \hbar ^\kappa
\left\vert D^{\kappa } 
p(\eta )(x) \right\vert_x ,
\end{equation} 
where $\vert \ \ \vert_x $ is a norm of covectors at $x \in M$.

In terms of the 
corresponding norm topology on $\Lambda^1 (M)$,\footnote{
Assume here that $\Lambda^1 (M)$ has the Banach
norm such that
$
\left\|  p  \right\| = \sup_{M} 
\sum_{\kappa \in {\bf Z}_{\geq 0} }\left\vert D^{\kappa } 
p (x) \right\vert_x ,
$
for $p\in \Lambda^1 (M)$.}
we can consider the space
$C^{\infty } \left( \Lambda^1 \left( M \right)   , C^{\infty }(M) \right) $
of all the $C^{\infty }$-differentiable
mapping from $\Lambda^1 \left( M \right) $ to $C^{\infty }(M)=C^{\infty }(M, {\bf R})$
and the subspaces
of the space
$ 
C( \Gamma  [
E ( M  )  ]   ) $ such that
\begin{equation}
C\left( \Gamma \left[
E\left( M \right) \right]  \right) 
= \left\{  \left. p^*F : 
\Gamma \left[
E ( M ) \right] \to C^{\infty }(M) \  \right\vert  
  F \in   
C^{\infty }\left( \Lambda^1  ( M  ) , C^{\infty }(M) \right)
 \right\} .
\end{equation}
Classical mechanics
requires the local 
dependence on
the momentum for functionals,
while quantum mechanics
needs the wider class
of functions
that depends on their derivatives.
The space of the classical
functionals
and
that of the quantum functionals
are defined as
\begin{eqnarray}
C_{cl}\left( \Gamma \left[
E\left( M \right) \right] \right) 
&=& \left\{ p^*F \in C\left( \Gamma \left[
E\left( M \right) \right] \right)
\ \left\vert \
 p^*F  \left( \eta \right) (x) = 
F^{T^*M} \left( x ,    p (\eta ) (x)  \right) \
\right.
\right\} \\
C_{q\ }\left( \Gamma \left[
E\left( M \right) \right]  \right) 
&=& \left\{   p^*F \in C \left( \Gamma \left[
E\left( M \right) \right] \right)
\ \right\vert \\
& \ & \ \ \ \ \ \ \ \ \ \ \left.
 p^*F  \left( \eta \right) (x) = 
F ^Q\left( x , p (\eta ) (x),  
..., D^{n} p (\eta ) (x) , ... \right) \ 
\right\} ,
\end{eqnarray}
and related with each other as
\begin{equation}\label{increasing}
C_{cl}\left( \Gamma \left[
E\left( M \right) \right]  \right) 
\subset
C_q\left( \Gamma \left[
E\left( M \right) \right]  \right) 
\subset
C\left( \Gamma \left[
E\left( M \right) \right] \right) .
\end{equation}
In other words,
the classical-limit
indicates
the limit of $\hbar \to 0$ with fixing $\vert p(\eta )(x) \vert $
finite at every $x\in M$, or what
the characteristic length $[x]$ and momentum $[p]$
such that $x/[x] \approx 1 $ and $p/[p] \approx 1 $
satisfies 
\begin{equation}
[p]^{-n-1} D^n p(\eta )(x)  \ll 1 .
\end{equation}
In addition,
the
$n$-th semi-classical system can
have the following functional space:
\begin{equation}
C_{n+1 }\left( \Gamma \left[
E\left( M \right) \right]  \right) 
= \left\{   p^*F \in C \left( \Gamma \left[
E\left( M \right) \right] \right)
\ \left\vert \
 p^*F  \left( \eta \right) (x) = 
F _{<n>}\left( x , p (\eta ) (x),  
..., D^{n} p (\eta ) (x) \right) \ 
\right.
\right\} .
\end{equation}
Thus, there is the increasing series
of subsets as
\begin{equation}
C_1 \left( \Gamma \left[ E(M)\right] \right)
... \subset  C_{n } \left( \Gamma \left[ E(M)\right] \right)
... \subset 
C_{\infty } \left( \Gamma \left[ E(M)\right] \right)
\subset 
C  \left( \Gamma \left[ E(M)\right]  \right) ,
\end{equation}
where $F _{<1>}=F ^{cl}$ and $F _{<\infty >}=F ^{q}$:
\begin{eqnarray}
C_{1\ } \left( \Gamma \left[ E(M)\right] \right)
&=& C_{cl} \left( \Gamma \left[ E(M)\right] \right) \\
C_{\infty } \left( \Gamma \left[ E(M)\right] \right)
&=&
C_{q \ } \left( \Gamma \left[ E(M)\right] \right) .
\end{eqnarray}

On the other hand,
the emergence-measure $\mu (\eta )  $ has
the Radon measure 
$ \tilde \mu (\eta )  $
for section $\eta \in \Gamma [E(M)]$ such that
\begin{equation}
\tilde \mu (\eta ) \ \left( F \left( p ( \eta  ) \right)
  \right)  =
\int_M d\mu (\eta ) (x) 
F \left( p ( \eta  ) \right) (x) .
\end{equation}
The introduced norm topology
on $ \Gamma \left( E(M) \right) $ induces the topological
$\sigma$-algebra
$ {\cal B}\left( {\cal O}_{ \Gamma \left( E(M) \right) }
\right) $;
thereby
manifold $ \Gamma \left( E(M) \right) $
becomes a measure space
having the probability 
measure ${\cal M}$ such that 
\begin{equation}
 {\cal M} \left(   
\Gamma \left( E(M) \right)  \right) =1.
\end{equation}
For a subset $C_n\left( \Gamma \left( E(M) \right)   \right)
\subset C\left( \Gamma \left( E(M) \right)   \right) $,
 an element
$\bar \mu 
\in C_n\left( \Gamma \left( E(M) \right)   \right) ^* $
is a linear functional
 $ \bar \mu : C_n\left( 
\Gamma \left[ E(M) \right] \right) 
 \to {\bf R}   $
such that
\begin{eqnarray}
 \bar  \mu \left( p^* F \right) &=&
\int_{\Gamma \left[ E(M) \right] }d {\cal M} (\eta ) \
\tilde  \mu (\eta ) \ \left( F \left( p ( \eta  ) \right)
  \right)  \\
&=& \int_{\Gamma \left[ E(M) \right]}d {\cal M} (\eta ) \
\int_M dv(x) \
\rho \left(\eta \right) (x)
F  \left( p ( \eta  ) \right) (x)   ,
\end{eqnarray}
where $ d \mu (\eta )   =
dv  \
\rho \left(\eta \right)   $.
Let us call mapping $\rho :\Gamma [E(M)]\to C^{\infty }(M)$
as the {\it emergence-density}.
The dual spaces
make an decreasing series
of subsets (consult \cite{Bogolubov} in the definition of the Gelfand triplet):
\begin{equation}
C_1 \left( \Gamma \left[ E(M)\right] \right) ^*
\supset ... 
C_n \left( \Gamma \left[ E(M)\right] \right) ^*
\supset ... 
C_{\infty } \left( \Gamma \left[ E(M)\right] \right) ^*\supset
C  \left( \Gamma \left[ E(M)\right]  \right) ^* .
\end{equation}
Thus,
relation
(\ref{increasing}) requires
the opposite sequence
for the dual spaces:
\begin{equation}\label{decreasing}
C_{cl}\left( \Gamma \left( E(M) \right)   \right) ^*
\supset
C_q\left( \Gamma \left( E(M) \right)   \right) ^*
\supset
C \left( \Gamma \left( E(M) \right)   \right) ^*.
\end{equation}
Let us summarize
how the relation between quantum mechanics and
classical mechanics in the following
diagram.

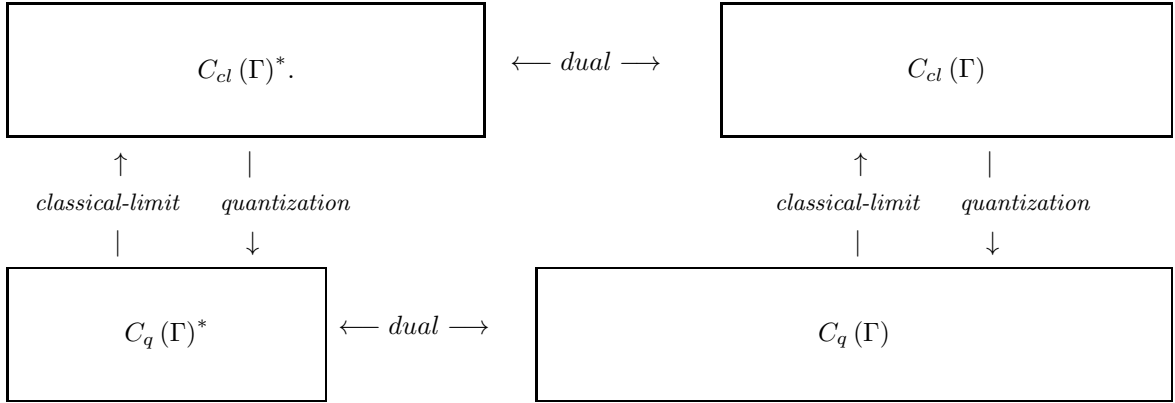
\begin{picture}(380,220)(10,20)
\put(210,50){\framebox(240,50){$C_{q}\left( \Gamma \right) $}}
\put(280,150){\framebox(170,50){$C_{cl}\left( \Gamma \right) $}}
\put(10,50){\framebox(120,50){$C_{q}\left( \Gamma \right) ^*$}}
\put(10,150){\framebox(180,50){$C_{cl}\left( \Gamma \right) ^*$.}}
\put(135,75){$ \longleftarrow   
dual  \longrightarrow $}
\put(200,175){$ \longleftarrow  
dual  \longrightarrow $}
\put(50,137){$ \uparrow $}
\put(20,122){{\small\it classical-limit}}
\put(50,107){$ \vert $}
\put(330,137){$ \uparrow $}
\put(300,122){{\small\it classical-limit}}
\put(330,107){$ \vert $}
\put(100,137){$ \vert $}
\put(90,122){{\small\it quantization}}
\put(100,107){$ \downarrow $}
\put(380,137){$ \vert $}
\put(370,122){{\small\it quantization}}
\put(380,107){$ \downarrow $}
\end{picture}

To investigate the time-development
of the statistical state discussed so far,
we will introduce the related group.
The group ${\cal D}(M)$ of
all the
$C^{\infty }$-diffeomorphisms of $M$
and the abelian group
$ C^{\infty }\left( 
M \right) $
 of all the $C^{\infty }$-functions 
on $M$
construct the semidirect product
$ S  ( M ) =
{\cal D} (M) \times_{semi. } C^{\infty }(M) $
of ${\cal D} (M)$ with $ C^{\infty }(M) $,
and define
the multiplication $\cdot $
between $ \Phi_1=(\varphi_1,s_1)$ and $\Phi_2=(\varphi_2,s_2)\in   S  ( M  )
$ as  
\begin{equation}\label{(2.1.1)}
\Phi_1\cdot \Phi_2 =(\varphi _1 \circ \varphi _2,(\varphi_2^*
s_1 )\cdot s_2) ,  
\end{equation}
for the pullback $\varphi ^*$ by $\varphi \in
{\cal D} (M)$.
The
Lie algebra 
$s (M)$
of $ S (M) $ 
has the Lie bracket such that, for   
$ V_1=(v_1 , U_1)$ and $V_2=(v_2 , U_2) \in  s (M)$,
\begin{equation}\label{(2.1.2)}
 [V_1, V_2 ]= \left( [v_1 ,v_2 ], v_1 U_2 - v_2 U_1 
+ \left[ U_1 , U_2 \right] \right) ;  
\end{equation}
 and its dual space $s (M)^*$
is defined by natural pairing $\langle \ , \  \rangle $.
Lie group
$ S  ( M  )  $ now acts on every $C^{\infty }$ section
 of $E ( M   ) $ (consult {\it APPENDIX}).
We shall further introduce the group $Q(M) 
= Map\left( \Gamma \left[ E(M)\right] , S  (M) \right) $ 
of all the mapping from $ \Gamma \left[ E(M)\right] $ into $S(M) $,
that has
the Lie algebra $q(M) 
= Map\left(  \Gamma \left[ E(M)\right] , s  (M)\right) $ and
its dual space $ 
q(M)^* = Map \left(   \Gamma \left[ E(M)\right] , s  (M)^* \right)  $.

Let us further define
the emergence-momentum
$ {\cal J}   
\in   q\left(M\right) ^*$ 
as follows:
\begin{equation}
 {\cal J}   \left( \eta  \right) 
=   d {\cal M} \left( \eta  \right) \
\left( \tilde    \mu \left( \eta  \right)  
 \otimes p (\eta )  ,\tilde    \mu \left( \eta  \right)  
\right)  .
\end{equation}
Thus,
the functional
${\cal F} : q\left(M\right) ^* \to {\bf R} $ 
can always be defined 
as 
\begin{equation}
{\cal F}  \left(  {\cal J} \right)
= \bar \mu  \left( p^*F \right) .
\end{equation}
On the other hand, the derivative
$  {\cal D}_{\rho  } 
F   \left( p \right)   $ can be 
introduced as follows
excepting the point
where the distribution $\rho $
becomes zero:
\begin{equation}
{\cal D}_{\rho  } 
F   \left( p \right)  (x)
=\sum_{ (n_1 ,  ... , n_N) \in {\bf N}^N}
{1\over { \rho  (x)} }
\left\{ \prod_{i}^{N} \left( - \partial_i \right) ^{ n_i }
  \left(  \rho  (x) 
p ( x )
{{\partial F    }\over { \partial \left\{
\left(
 \prod_{i}^{N}
\partial_i^{n_i } \right) p_j \right\}  }}
\right)  \right\}
\partial_j .
\end{equation}
Then,
operator $\hat F \left( \eta \right)  =
{{\partial {\cal F}}
\over {\partial {\cal J}}}  \left( {\cal J}\left( \eta \right)\right) $ 
is defined as
\begin{equation}
\left. {d\over d\epsilon }\right\vert_{\epsilon =0}
{\cal F } \left( {\cal J}  + \epsilon {\cal K}\right) 
= \left\langle  {\cal K} , \hat F   \right\rangle ,
\end{equation}
i.e.,
\begin{equation}
\hat F  \left( \eta \right)
= \left(   {\cal D}_{ \rho (\eta ) } F  
\left(  p  (\eta ) \right) ,
- p  (\eta )  \cdot     {\cal D}_{ \rho (\eta ) } F   
\left(  p  (\eta ) \right)
+ F    \left( p  (\eta )  \right)
\right)  ;
\end{equation}
thereby, the following null-lagrangian relation
 can be obtained:
\begin{equation}
{\cal F}  
 \left( {\cal J}   \right) 
= 
\langle {\cal J}   , \hat F    \rangle  .
\end{equation}

Let us consider the time-development
of the section $\eta ^{\tau }_{t }(\eta ) \in \Gamma [E(M)]$
such that
 the {\it labeling time} $\tau $ satisfies
$\eta ^{\tau }_{\tau }(\eta ) =\eta  $.
It has the momentum
 $p_t^{\tau }(\eta )
=-i \bar h\eta ^{\tau }_t(\eta )^{-1}
d \eta ^{\tau }_t(\eta ) $ and
the emergence-measure $\mu^{\tau }_t(\eta ) $
such that
\begin{equation}\label{q-measure-rel}
d {\cal M} \left( \eta \right) \
  \tilde \mu_t^{\tau } \left( \eta  \right) 
= d {\cal M} \left(  \eta ^{\tau }_t(\eta )\right) \
\tilde \mu_t \left( \eta ^{\tau }_t(\eta )\right) :
\end{equation}
\begin{eqnarray}\label{mu(F)}
\bar  \mu_t \left( p^*F_t\right) 
&=&\int_{\Gamma \left[ E(M) \right]}d {\cal M} (\eta ) \
 \tilde \mu_t (\eta ) \ \left( p^*F_t  (\eta )
  \right)  \\
&=& \int_{\Gamma \left[ E(M) \right]}d {\cal M} \left( \eta \right) \
\tilde \mu^{\tau }_t \left( \eta  \right) \ \left( p^*F  \left( 
\eta ^{\tau }_t(\eta )\right) 
  \right)  \\
&=&
 \int_{\Gamma \left[ E(M) \right]}d {\cal M} \left( \eta \right) \
\int_M dv(x) \ \rho_t^{\tau } (\eta ) (x) 
F_t   \left( p ^{\tau }_t(\eta )\right)  (x)    .
\end{eqnarray}
The introduced labeling time $\tau $
can always be chosen such that $ \eta ^{\tau }_t(\eta ) $
does not have any singularity
within a short time for every $\eta
\in \Gamma \left[ E(M) \right]$.
The emergence-momentum
$ {\cal J}_t^{\tau }  
\in   q\left(M\right) ^*$ 
such that
\begin{eqnarray}
{\cal J}_t^{\tau }  (\eta ) &=&
 {\cal J}_t  \left( \eta ^{\tau }_t(\eta )\right)  \\
&=&   d {\cal M} \left( \eta ^{\tau }_t(\eta )\right) \
\left( \tilde   \mu_t \left( \eta ^{\tau }_t(\eta )\right)  
 \otimes 
p_t^{\tau }(\eta )  ,\tilde   \mu_t \left( \eta ^{\tau }_t(\eta )\right)  \right) \\
&=&  d {\cal M} (\eta ) \
 \left(  \tilde \mu_t ^{\tau }
\left( \eta  \right) \otimes 
p_t^{\tau }(\eta )  ,
\tilde \mu_t ^{\tau }
\left( \eta  \right) \right) \ 
\end{eqnarray}
satisfies the following relation
for the functional
${\cal F}_t   : q\left(M\right) ^* \to {\bf R} $:
\begin{equation}
{\cal F}_t  \left(  {\cal J}^{\tau }_t\right)
=  \mu_t \left( p^* F_t\right) ,
\end{equation}
whose value is independent of
labeling time $\tau $.
The operator $\hat F_t^{\tau } =
{{\partial {\cal F}_t}
\over {\partial {\cal J}}}  \left( {\cal J}^{\tau }_t\right) $ 
is defined as
\begin{equation}
\left. {d\over d\epsilon }\right\vert_{\epsilon =0}
{\cal F }_t  \left( {\cal J}^{\tau }_t + \epsilon {\cal K}\right) 
= \left\langle  {\cal K} , \hat F_t^{\tau }  \right\rangle ,
\end{equation}
i.e.,
\begin{equation}
\hat F_t  ^{\tau }
= \left(   {\cal D}_{ \rho_t^{\tau }(\eta ) } F_t  
\left(  p^{\tau }_t (\eta ) \right) ,
- p_t^{\tau } (\eta )  \cdot     {\cal D}_{ \rho_t^{\tau }(\eta ) } F_t  
\left(  p^{\tau }_t (\eta ) \right)
+ F_t   \left( p^{\tau }_t (\eta )  \right)
\right)  .
\end{equation}
Thus, the following null-lagrangian relation
 can be obtained:
\begin{equation}
{\cal F}_t   
 \left( {\cal J}_t^{\tau }  \right) 
= 
\langle {\cal J}_t^{\tau }  , \hat F _t ^{\tau } \rangle  ,
\end{equation}
while
the normalization condition 
has the following expression:
\begin{equation}\label{q-normalize}
{\cal I}\left( {\cal J}^{\tau }_t \right) =1 
\ \ \ \ \ for \ \ \ \ \ 
{\cal I}\left( {\cal J}^{\tau }_t \right) =
\int_{\Gamma \left[ E(M) \right] }d{\cal M}(\eta )  \
\mu_t (\eta ) (M) .
\end{equation}

\begin{tm}
For Hamiltonian operator $\hat H_t^{\tau }  =
{{\partial {\cal H}_t}\over {\partial {\cal J}}} 
  \left( {\cal J}_t^{\tau } \right)\in   q\left(  M\right) $
corresponding to Hamiltonian
$p^*H_t \left( \eta \right) (x) =
H^{T^*M}_t\left( x, p \left( \eta \right) \right) $,
equations (\ref{internal Lie velocity}) and (\ref{quotient}) of motion
becomes Lie-Poisson equation 
\begin{equation}\label{Hq}
{{\partial  {\cal J} ^{\tau }_t}\over {\partial t}} 
= ad^*_{\hat H_t ^{\tau }     }{\cal J}^{\tau } _t   ,
\end{equation}
which can be expressed as
\begin{equation}
{{\partial  } \over {\partial t}} 
 \rho _t^{\tau }(\eta )(x) 
= -\surd ^{-1}\partial_j \left( 
{ { \partial H_t^{T^*M}  }\over {\partial \ p_j \  }} 
\left( x, p_t^{\tau } \left( \eta \right) (x) \right) 
 \rho  _t^{\tau }(\eta )(x) \surd  \right) ,
\end{equation}
\begin{eqnarray}\nonumber
{{\partial  } \over {\partial t}}
\left( \rho  _t^{\tau }(\eta )(x) 
  p_{tk}^{\tau } (\eta ) (x) \right)
\nonumber
&=&
 - \surd^{-1} \partial_j \left(
{ { \partial H_t^{T^*M}  }\over {\partial \ p_j \  }} 
\left( x, p_t^{\tau } \left( \eta \right) (x) \right) 
\rho  _t^{\tau }(\eta )(x)  p_{tk}^{\tau } (\eta )(x) \surd  
\right) \\
\nonumber
&  &
-  \rho  _t^{\tau }(\eta )(x)  p_{tj}^{\tau } (\eta ) (x) 
\partial_k  
{ { \partial H_t^{T^*M}  }\over {\partial \ p_j \  }} 
\left( x, p_t^{\tau } \left( \eta \right) (x) \right)   \\
&  &
+ \rho  _t^{\tau }(\eta ) (x)  
\partial_k \left(
 p^{\tau }_t (\eta ) (x) \cdot 
{ { \partial H_t^{T^*M}  }\over {\partial \ p \  }} 
\left( x, p_t^{\tau } \left( \eta \right) (x) \right) 
\right. \\
&  & \ \ \ \ \ \ \ \ \ \ \ \ \ \ \ \ \
\left.
- H_t^{T^*M}   \left( x, p^{\tau }_t (\eta ) (x) \right) 
\right)  .
\end{eqnarray}
\end{tm}
\par\noindent${\it Proof.}\ \ $
{\it
Lie-Poisson equation (\ref{Hq})
 is  calculated 
for ${\cal D} H_t  ^{\tau }(\eta ) 
=  {\cal D}_{ \rho_t^{\tau }(\eta ) } H_t  
\left(  p^{\tau }_t (\eta ) \right) $
as follows:
\begin{equation}
\label{density's-q} 
{{\partial  } \over {\partial t}} 
 \rho _t^{\tau }(\eta )(x) 
= -\surd ^{-1}\partial_j \left(
{\cal D}^j H_t  ^{\tau }(\eta ) (x)
 \rho  _t^{\tau }(\eta )(x) \surd  \right) ,
\end{equation}
\begin{eqnarray}\nonumber
{{\partial  } \over {\partial t}}
\left( \rho  _t^{\tau }(\eta )(x) 
  p_{tk}^{\tau } (\eta ) (x) \right)
\nonumber
&=&
 - \surd^{-1} \partial_j \left(
{\cal D}^j H_t  ^{\tau }(\eta ) (x)
\rho  _t^{\tau }(\eta )(x)  p_{tk}^{\tau } (\eta )(x) \surd  
\right) \\
\nonumber
&  &
-  \rho  _t^{\tau }(\eta )(x)  p_{tj}^{\tau } (\eta ) (x) 
\partial_k  
{\cal D}^j H_t  ^{\tau }(\eta ) 
 (x)  \\
\label{current's-q}
&  &
+ \rho  _t^{\tau }(\eta ) (x)  
\partial_k \left(
 p^{\tau }_t (\eta ) (x) \cdot {\cal D} H_t  ^{\tau }(\eta )  (x)
- H_t   \left( p^{\tau }_t (\eta ) \right) (x)
\right)  ,
\end{eqnarray}
where $ dv= dx^1 \wedge ... dx^N \ \surd $
and
$\surd = \sqrt{det \left\vert g^{jk} \right\vert }$ 
for the
local coordinate ${\bf x}= \left( x^1, x^2, ... , x^N\right) $.
Second equation (\ref{current's-q})
can be rewritten in conjunction with 
the conservation  (\ref{density's-q}) of the
emergence-density
as
\begin{equation}
{{\partial  } \over {\partial t}}
  p_{tk}^{\tau } (\eta ) (x) 
+   
{\cal D}^j H_t  ^{\tau }(\eta ) (x)
 \partial_j   p_{tk}^{\tau } (\eta )(x)  
+     p_{tj}^{\tau } (\eta ) (x) 
\partial_k  
{\cal D}^j H_t  ^{\tau }(\eta ) 
 (x) =  
\partial_k  L^{\tau }_t (\eta ) (x) ,
\end{equation}
where
\begin{equation}
 L^{\tau }_t (\eta ) (x)= p^{\tau }_t (\eta )
 (x) \cdot {\cal D} H_t  ^{\tau }(\eta )  (x)
- H_t   \left( p^{\tau }_t (\eta ) \right) (x) ,
\end{equation}
or, by using Lie derivatives,
\begin{equation}
{\cal L}_{{\cal D} H_t  ^{\tau }(\eta )  }
\  p_{t}^{\tau } (\eta )  =  
d L^{\tau }_t (\eta ) .
\end{equation}
Thus,
we can obtain the 
equation of motion
in the following simpler form
by using Lie derivatives:
\begin{eqnarray}
{\cal L}_{{\cal D} H_t  ^{\tau }(\eta )  }
\  \eta _{t}^{\tau }    &=&  -i\bar h
 L^{\tau }_t (\eta )  \  \eta _{t}^{\tau } \\
{\cal L}_{{\cal D} H_t  ^{\tau }(\eta )  }
\ \rho _t^{\tau }(\eta ) \ dv 
&=& 0 ,
\end{eqnarray}
which is equivalent to the equations  (\ref{internal Lie velocity}) and (\ref{quotient}) 
when $p^*H_t \left( \eta \right) (x) =
H^{T^*M}_t\left( x, p \left( \eta \right) \right) $

}\hspace{\fill} { \fbox {}}\\

\noindent
Equation (\ref{Hq}) will
prove in the following two sections
to include the
Schr\"odinger equation 
in canonical quantum mechanics
and the classical Liouville equations
in classical mechanics.

For
$  {\cal U}_t  ^{\tau }
 \in  Q\left(  M\right) $ such that
$
{{\partial  
{\cal U}_t  ^{\tau } }\over {\partial t}}  
\circ \left( {\cal U}_t  ^{\tau  } \right) ^{-1}= 
\hat H_t  ^{\tau } (\eta ) 
\in q(M)  $,
let us introduce the following operators:
\begin{eqnarray}
  \tilde H_t^{\tau } (\eta )   = Ad^{-1}_{{\cal U}_t  ^{\tau }   }
\hat H_t^{\tau } (\eta )  \ \ \left( =  \hat H_t ^{\tau } (\eta )\right) , \   and \ \ \
 \tilde F_t^{\tau } (\eta )  = Ad^{-1}_{{\cal U}_t  ^{\tau }    }
\hat F_t ^{\tau } (\eta ) .
\end{eqnarray}
It satisfies the following theorem.

\begin{tm}
Lie-Poisson equation (\ref{Hq})
is
equivalent to the following equation:
\begin{equation}\label{general heisenberg}
{{\partial }\over {\partial t}}\tilde F_t  ^{\tau }    = 
\left[ \tilde H _t ^{\tau }    ,\tilde F_t   ^{\tau }   \right] 
+\widetilde{\left( {{\partial F_t  ^{\tau }     }
\over {\partial t}} \right) } .
\end{equation}
\end{tm}
\par\noindent${\it Proof.}\ \ $
{\it
Equation (\ref{Hq}) of motion
concludes the following equation:
\begin{equation}
\left\langle
{{\partial  {\cal J} ^{\tau }_t}\over {\partial t}}  ,\hat F_t   ^{\tau }  
\right\rangle
= \left\langle
ad^*_{\hat H_t ^{\tau }     }{\cal J}^{\tau } _t  ,
\hat F_t   ^{\tau }  
\right\rangle .
\end{equation}
The left hand side can be calculated as
\begin{eqnarray}
L.H.S.
&= & 
{{d  }\over {d t}} {\cal F}_t\left( {\cal J}^{\tau }_t\right)
-{{\partial  {\cal F}_t}\over {\partial t}} \left( {\cal J}^{\tau }_t\right)\\
&= &
\left\langle \left(
{{\partial  }\over {\partial t}}Ad^*_{{\cal U}_t^{\tau}}{\cal J} ^{\tau }_{\tau }  
\right) ,
\hat F_t   ^{\tau }  
\right\rangle  -
\left\langle
Ad^*_{{\cal U}_t^{\tau}}{\cal J} ^{\tau }_{\tau }  ,
\hat {{\partial  F_t^{\tau }}\over {\partial t}} 
\right\rangle \\
&=&
\left\langle
{\cal J} ^{\tau }_{\tau }  ,
{{\partial  }\over {\partial t}} \tilde F_t   ^{\tau }  
\right\rangle
 -
\left\langle
{\cal J} ^{\tau }_{\tau }  ,
\tilde {{\partial  F_t^{\tau }}\over {\partial t}} 
\right\rangle  ;
\end{eqnarray}
and the right hand side becomes
\begin{eqnarray}
R.H.S.
&= & 
\left\langle
ad^*_{\hat H_t ^{\tau }     }Ad^*_{{\cal U}_t ^{\tau }}{\cal J}^{\tau } _t  ,
\hat F_t   ^{\tau }  
\right\rangle \\
&=&
\left\langle
Ad^*_{{\cal U}_t ^{\tau }}ad^*_{\tilde H_t ^{\tau }     }{\cal J}^{\tau } _t  ,
\hat F_t   ^{\tau }  
\right\rangle \\
&=&\left\langle
{\cal J}^{\tau } _t  , \left[
\tilde H_t ^{\tau }    ,
\tilde F_t   ^{\tau }  \right]  
\right\rangle .
\end{eqnarray}
Thus, we can obtain this theorem.
}\hspace{\fill} { \fbox {}}\\

The general theory
for Lie-Poisson systems
certificates that,
 if a  group action of Lie group
$ Q(M)$
keeps the Hamiltonian ${\cal H}_t: q(M)^* \to {\bf R}$  
invariant,
there exists an invariant charge 
functional
$Q : \Gamma \left[ E(M)\right] \to
C(M) $ and the induced
function ${\cal Q}: q(M)^* \to {\bf R}$ such that
\begin{equation}\label{charge-invariant}
\left[ \hat H _t, \hat Q \right]  = 0 ,
\end{equation}
where $\hat Q $ is expressed as
\begin{equation}
\hat Q 
= \left(  {\cal D}_{\rho (\eta )} Q
\left(  p  (\eta ) \right) ,
- p  (\eta )  \cdot     {\cal D}_{\rho (\eta )} Q
\left(  p  (\eta ) \right)
+ Q   \left( p (\eta )  \right)
\right)  .
\end{equation}

\section{CONCLUSION}
The present paper
attempted to reveal the structure behind mechanics,
and proposed a basic theory of
physical reality realizing Whitehead's philosophy.
It induced 
protomechanics that
deepened Hamiltonian mechanics
under the modified  Einstein-de Broglie relation.
In the following papers \cite{SbMII,SbMIII},
the present theory will
prove to induce both classical mechanics
and quantum mechanics,
to solve the problem of the operator ordering
in quantum mechanics and to
give its realistic, 
self-consistent interpretation.

\setcounter{equation}{0}
\renewcommand{\theequation}{A\arabic{equation}}
\section*{APPENDIX: LIE-POISSON MECHANICS}\label{Lie-Poisson Mechanics}

Over a century ago, 
in an effort to elucidate the relationship 
between Lie group theory and classical mechanics,
Lie \cite{Lie} introduced the
 {\it Lie-Poisson system}, being
a Hamiltonian system 
on the dual space of an arbitrary finite-dimensional Lie algebra.
Several years later, as a generalization of
the
Euler equation of a rigid body, Poincar\'e \cite{Poincare} 
applied the standard variational principle on the  
tangent space of an arbitrary finite-dimensional Lie group and
independently
obtained the {\it Euler-Poincar\'e equation}
on the Lie algebra,
being
equivalent to
the {\it Lie-Poisson equation} on 
its dual space if considering no analytical difficulties.
These mechanics structures
for Lie groups
were reconsidered 
in the 1960's
(see  \cite{M&R} for the historical information).
Marsden and Weinstein \cite{M&W2}, in 1974,
proposed the 
{\it Marsden-Weinstein reduction method}  
that allows a Hamiltonian system
to be reduced due to
the symmetry determined by an appropriate
Lie group, while
Guillemin and Sternberg \cite{G&S}
introduced
{\it the collective-Hamiltonian method}  that
describes the equation of motion for a Hamiltonian system as
the Lie-Poisson equation of a reduced Lie-Poisson system.

Let 
$G$ be taken to be  a  finite-  or infinite-dimensional Lie group
and ${  g}$ 
the Lie algebra of $G$; i.e.,
the multiplications $ \ \ \cdot \ \  : G \times G \to G:
(\phi _1 , \phi _2 )\to \phi _1 \cdot  \phi _2 $ 
with a unit $e \in G$
satisfy $\phi _1^{-1} \cdot \phi _2 \in G$
and induce
the commutation relation
$[ \ \ , \ \  ] :  {  g}\times  {  g} \to  {  g} : ( v_1 , v_2 )
\to [ v_1 , v_2]  $.
For a function
$F \in C^{\infty }( G , {\bf R}) $,
two types of derivatives  respectively 
define  the left- and the right-invariant vector
field $v^+ $ and $v^- \in {\cal X}(G)$
in the space ${\cal X}(G)$ of all smooth vector fields
on $G$: 
\begin{eqnarray}
v^+  F(\phi) &=& {d \over  d \tau } \vert_{\tau =0}
F( \phi \cdot  e^{\tau v})
\\
v^-  F(\phi) &=& {d \over  d \tau } \vert_{\tau =0}
F( e^{\tau v} \cdot  \phi )  .
\end{eqnarray}
Accordingly, the left- and  the right-invariant element
of the space ${\cal X}(G)$  satisfy 
\begin{equation}
[ v_1^+
, v_2 ^+]= [ v_1 , v_2]^+ 
,\ \ \
[ v_1^-, v_2^- ]=- [ v_1 , v_2]^- 
,\ \ \ 
and \ \ \ \ [ v_1^+, v_2^- ]=0 .
\end{equation} 
In the subsequent formulation,
$+$ and $-$ denote 
left- and right-invariance, respectively.
In addition, $\langle   \ \   ,  \ \  \rangle :
{  g}^* \times {  g} \to {\bf R}:
(\mu  , v )\to \langle \mu   , v \rangle $ denotes 
the  nondegenerate natural pairing (that is weak in general
\cite{A&M}) for the dual space ${  g}^*$ of the Lie algebra ${ 
g}$, defining 
the left- or right-invariant 1-form
$\mu ^{\pm } \in  \Lambda^1 (G) $ corresponding to
$\mu \in {  g}^*$
by introducing the 
natural pairing  $\langle   \ \   ,  \ \  \rangle :
T^*_{\phi }G \times T_{\phi } G \to {\bf R}$
for $\phi \in G$
as
\begin{equation}
  \langle \mu^{\pm } (\phi ), v^{\pm } (\phi ) \rangle 
= \langle  \mu  , v \rangle .
\end{equation}

Let us now consider
how the motion
on a
Poisson manifold $P$
can be represented by the Lie-Poisson 
equation for $G$ (or its central extension \cite{A&M}),
where $P$ is a finite or infinite Poisson manifold
modeled on $C^{\infty }$ Banach spaces
with Poisson bracket $\{  \ \ , \ \  \} :
C^{\infty }(P, {\bf R})\times C^{\infty }(P, {\bf R}) 
\to C^{\infty }(P, {\bf R})$.
Also, $\Psi :G \times P
\to P$ is an action of $G$ on $P$
such that
the mapping $\Psi_{\phi } : P \to P$ is a
Poisson mapping
for each $\phi  \in G$
in which
$\Psi_{\phi }(y)
= \Psi (\phi , y)$ for $ y \in P $.
It is assumed that 
the Hamiltonian mapping $\hat J : {  g} \to C^{\infty }(P, {\bf R})$
is obtained
for this action
s.t. $X_{\hat J (v)}=v_P $
for $ v  \in {  g}$,
where $X_{\hat J (v)}$
and $v _P \in {\cal X}(P)$ denote 
the Hamiltonian vector field for $\hat J (v) \in C^{\infty }(P, {\bf R})$
and
the infinitesimal generator of 
the action on $P$ corresponding to
$v \in {  g} $, respectively.
As such, the momentum (moment) mapping 
$J : P \to {  g}^*$
is defined 
by $\hat J (v) (y) = \langle J(y), v \rangle $. 
 For 
the special case in which 
$(P,\omega )$ is a
symplectic manifold 
with a symplectic 2-form  $\omega \in \Lambda^2 (G)$ 
(i.e., $d \omega =0 $ and $\omega $ is weak nondegenerate),
this momentum mapping 
is equivalent to that defined by 
$ d\hat J ( v ) = v _P \rfloor \omega $.
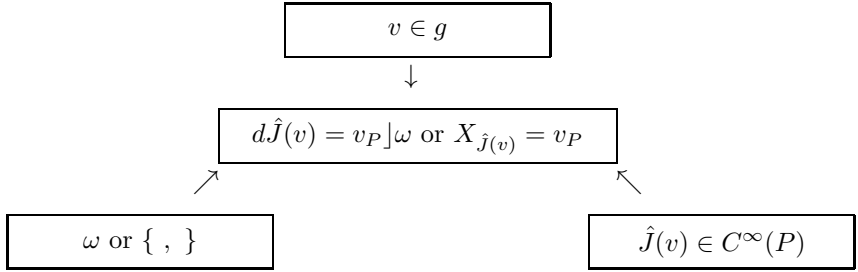
$$ $$
\begin{picture}(400,100)(0,0)
\put (185,80){\framebox(100,20){$v \in {  g} $}}
\put (150,30){$\nearrow $}
\put (310,30){$\nwarrow $}
\put (230,70){$\downarrow $}
\put (300,0){\framebox(100,20){$\hat J (v) \in C^{\infty }(P) $}}
\put (80,0){\framebox(100,20){$\omega $ or
$\{ \ , \ \} $}}
\put (160,40){\framebox(150,20){$ d\hat J ( v ) = v _P \rfloor
\omega $ or $ X_{\hat J
( v )} = v _P $}}
\end{picture}
$$ $$
In twentieth century, 
lots of mathematicians would have based their study
especially on the Poisson structure or the symplectic structure
in
the above diagram,
while the physicists 
would usually have made importance the functions
as the Hamiltonian and the other invariance of motions
as some physical matter.
In  Lie-Poisson mechanics,
the Lie group plays the most important
role as "motion" itself,
while the present theory inherits such an idea.

For the trivial topology
of $G$ (consult \cite{A&M}
in the nontrivial cases),
the Poisson bracket satisfies
\begin{equation} \label{one-cocycle}
\{\hat J (v_1), \hat J(v_2) \} = \pm \hat J([v_1,v_2] )  .
\end{equation}
The {\it Collective Hamiltonian Theorem} \cite{M&R}
concludes
the Poisson bracket for $ A\circ J $ and $ B\circ J
\in C^{\infty }(P, {\bf R})$ 
can be expressed for $\mu = J(y) \in {  g}^*$
 as
\begin{equation} \label{Poisson}
\{ A \circ J , B \circ J \} (y) =   \pm \langle J(y) , 
[{{\partial A }\over {\partial \mu  }}(\mu ),
{{\partial B }\over {\partial
\mu  }}(\mu )] \rangle  ,
\end{equation}
where 
$ 
{{\partial F }\over {\partial \mu  }} :
{  g}^* \to {  g}
$ is the Fr\'echet derivative of 
$ F \in C^{\infty }({  g}^*, {\bf R})$
that  every $\mu \in {  g}^*$ and $ \xi \in {  g}$ satisfies
\begin{equation}
{d \over {d \tau }}\vert_{\tau =0 } F(\mu + \tau \xi )
= \left\langle \xi , {{\partial F }\over {\partial \mu }} (\mu )
\right\rangle .
\end{equation}
Thus,
the collective 
Hamiltonian $H\in  
C^{\infty }( {  g}, {\bf R}) $ such that $ H_P=H\circ J $ 
collects or reduces
 the   Poisson  equation of motion
into the following Lie-Poisson  equation of motion:
\begin{equation} \label{ad*eq}
 {d \over {d t }}\mu_t = 
\pm ad^*_{{{\partial H }\over {\partial  \mu }}(\mu_t )} 
\mu_t ,
\end{equation}
where $ \mu_t = J  (x_t) $ for $x_t \in P$.
We can further obtain the formal solution of Lie-Poisson equation of motion
(\ref{ad*eq}) as
\begin{equation}
 \mu_t   = Ad^*_{\phi_t }\mu_0 ,
\end{equation}
where
generator $\phi_t \in \tilde G$ satisfies
$\{ {{\partial H}\over {\partial \mu }}(\mu_t) \}^{+}
= \phi_t ^{-1}\cdot {{d \phi_t }\over {dt}}  $
or
$\{ {{\partial H}\over {\partial \mu }}(\mu_t) \}^{-}
= {{d \phi_t }\over {dt}}  \cdot \phi_t ^{-1}$
The existence of this solution, however, 
should independently verified
(see \cite{E&M} for example).

In particular,
Arnold \cite{Arnold}
applies such group-theoretic method not only to
the equations of motion of a rigid body
but also to
that of an ideal incompressible fluid,
and constructs them as the motion of a particle
on the three-dimensional special 
orthogonal group $SO(3)$ and as that on the infinite-dimensional Lie group 
${\cal D}_v(M)$ of
all $C^{\infty }$ volume-preserving diffeomorphisms
on a compact oriented manifold $M$.
By introducing semidirect products of Lie algebras,
Holm and Kupershmidt \cite{H&K}
and
Marsden {\it et al.} 
\cite{M&R&W} went on to 
complete the method
such that various Hamiltonian systems can be 
treated as Lie-Poisson systems, e.g.,
the motion of a top under gravity and
that of an ideal magnetohydodynamics (MHD) fluid.

For the motion of an isentropic fluid,
the governing Lie group is a semidirect product of 
the Lie group ${\cal D}
 (M)$
of all $C^{\infty }$-diffeomorphisms
on $M$
with 
$C^{\infty }(M)\times C^{\infty }(M)$, i.e., 
\begin{equation}\label{(3.1)}
G(M)={\cal D}(M)\times_{semi.} \left\{ C^{\infty }(M)\times  C^{\infty }(M) 
\right\} .
\end{equation}
For $ \tilde \phi_1=(\phi_1, f_1, g_1)$,
$ \tilde \phi_2=(\phi_2,  f_2, g_2)\in I(M)$, 
the product of two elements of  $I(M)$ is
defined as follows: 
\begin{eqnarray} \nonumber
  \tilde \phi_1 \cdot  \tilde \phi_2 &=&  (\phi_1, f_1,g_1)
\cdot (\phi_2,  f_2,g_1)
\\ \label{(3.2)}
&=&  \left( \phi_1\circ \phi
_2,
\phi_2^* f_1+f_2 ,\phi_2^*
g_1+g_2 
\right) 
\ , 
\end{eqnarray}     
where $\phi^*$ denotes the pullback by $\phi\in
{\cal D}(M)$ and 
the unit element of $G(M)$ can be denoted as
$(id. ,0, 0) \in G(M)$, where $id. 
\in {\cal D} (M)$ is the identity mapping from $M$ to itself. 

The
Lie bracket 
for $\tilde v_1 
=( v_1^i\partial
_i, U_1, W_1 ) 
$ and $\tilde v_2 =\left( v_2^i\partial _i ,   U_2, 
W_2 \right)  \in g(M)$ becomes 
\begin{equation}
\left[ \tilde v_1^-, \tilde v_2^-\right] 
=\left( 
 \left[ {v_1^i\partial _i,
v_2^j\partial _j } \right]
,  
v_1^j\partial _jU_2 -
v_2^j\partial _jU_1 ,  
 v_1^j\partial _jW_2 -
v_2^j\partial _jW_1  
\right)  .
\end{equation}
For the volume measure $v$ of $M$,
the element 
of the dual
space $g (M)^*$ 
of the Lie algebra $g (M)$
can
be described  as
\begin{equation}
{\cal J}_t  = \left( dv \ \rho_t \otimes p _t ,
 dv \ \rho_t ,  dv \ \sigma_t \right) ,
\end{equation}
in that $ p _t \in \Lambda^1(M) $,
$ dv \ \rho_t \in \Lambda^3(M)$
 and $ dv \ \sigma_t \in \Lambda^3(M)$
 physically means the 
momentum, the
mass density, and the entropy density.

For the thermodynamic internal energy $ U\left( \rho(x),
\sigma (x)\right) $,
the 
 Hamiltonian 
for the motion of an
isentropic fluid
is introduced as
\begin{equation}
 {\cal H}\left( {\cal J}  \right) =
{1 \over 2}  \int_M dv(x) \ \rho_t (x)  g^{ij}(x) p _{t j} p _{t j}
+
 \int_M dv(x) \ \rho_t (x)  U(\rho_t(x), \sigma_t
(x))     .
\end{equation}
Define
the
operator $\hat F_t =
{{\partial {\cal F} }
\over {\partial {\cal J}}}  \left( {\cal J} _t \right) 
\in g (M)$ 
for every functional $F : g (M)^*
\to {\bf R}$
as
\begin{equation}
\left. {d\over d\epsilon }\right\vert_{\epsilon =0}
{\cal F } \left( {\cal J} _t + \epsilon {\cal K}\right) 
= \left\langle  {\cal K} , \hat F_t  \right\rangle ,
\end{equation}
then,
the Hamiltonian operator $\hat H_t  =
{{\partial {\cal H}_t}\over {\partial {\cal J}}} 
  ({\cal J} _t )\in g(M) $
is calculated for the velocity field $v_t =g^{ij}p_i  \partial_j \in X^1(M)$
as
\begin{equation}
\hat H_t = \left( v^j \partial_j  , 
-{1\over 2}g^{ij }p_{ti} p_{tj} + U\left( \rho_t (x) , \sigma_t (x) \right)
+ \rho_t (x) {{\partial U}\over {\partial \rho }}\left( \rho_t (x) , \sigma_t (x) \right)
,
\rho_t (x) {{\partial U}\over {\partial \sigma }}\left( \rho_t (x) , \sigma_t (x) \right)
\right) .
\end{equation}
The equation of motion
becomes the following Lie-Poisson equation:
\begin{equation}\label{compr.LP}
{{d {\cal J} _t}\over {d t}} 
= ad^*_{\hat H_t  }{\cal J} _t   ,
\end{equation}
which  is  calculated 
as follows:
\begin{enumerate}
\item the conservation laws of
mass and entropy:
\begin{eqnarray}
{{\partial
\bar \rho_t } \over {\partial t }}+  \surd^{-1} 
\partial_j \left( \rho_t v^j _t  \surd \right) &=&0  ,\\
 {{\partial
\bar \sigma_t } \over {\partial t }}+ 
 \surd^{-1} 
\partial_j \left( \sigma_t v^j _t  \surd \right)
 &=&0  ,
\end{eqnarray}
where $\surd = \sqrt{\vert det g^{ij }\vert }$;
\item the conservation law of
momentum:
\begin{equation}
{{\partial  } \over {\partial t}}
\left( \rho  _t  
  p_{tk}    \right)
+
 \surd^{-1} \partial_j \left(
v^j  
\rho  _t    p_{tk}   \surd  
\right)   +  \partial_k P_t =0 ,
\end{equation}
where the pressure $ P_t $ satisfies the 
following condition:
\begin{equation}\label{(3.23)}
 P_t (x) = \rho_t (x) \{
\rho_t (x) { { \partial U } \over {
\partial \rho }}  +\sigma_t (x) { { \partial U } \over
{ \partial \sigma }} \} \left( \rho_t (x) , \sigma_t (x) \right) , 
\end{equation}
which is consistent with the first law of thermodynamics.
\end{enumerate}

Next, we consider ${\cal D}_{v}(M)  $, being the Lie group of
volume-preserving diffeomorphisms of $M$,
where
every element $\phi \in {\cal D}_{v}(M) $ satisfies 
$dv \left( \phi (x) \right) =dv (x)$.
Lie group ${\cal D}_{v}(M)$ is  a subgroup of  $G(M) $,
and inherits its Lie-algebraic structure of.
A right-invariant vector at $T_e{\cal D}_{v}(M)$
is identified with the corresponding divergence-free
vector field on $M$, i.e.,    
\begin{equation}\label{(2.14)}
u^- (e)=u^i\partial _i
\quad \quad \nabla \cdot {\bf u}=0 \quad for \quad all
\quad x \in M .
\end{equation} 
We can define an operator
$P_{\phi}$ \cite{E&M} that orthogonally projects the elements of
$T_{\phi}G(M)$ onto $T_{\phi}{\cal D}_v(M)$ for $\phi \in {\cal D}_v(M)
 \subset G(M)$
such that
\begin{equation}\label{(2.15)}
P_{\phi}[v^- (\phi )]=P[v]^- (\phi )
\end{equation}
and
\begin{equation}\label{(2.16)}
 P[v]^- (e)  =
(v^i - \partial^i \theta )\partial _i, 
\end{equation}
where  $\theta : M \to {\bf R}$ satisfies
$\partial _i(v^i(x) - \partial^i \theta (x))=0$
for every $x \in M$.
This projection
changes   
Lie Poisson equation
(\ref{compr.LP})
into the new Lie-Poisson equation
representing
the Euler equation for the motion
of an incompressible fluid:
\begin{equation}
\quad{{\partial {\bf u}_t} \over {\partial t}}+{\bf
u}_t\cdot \nabla {\bf u}_t+\nabla p=0, 
\end{equation}    
where the pressure $p : M \to {\bf R}$ is 
determined by the condition $\nabla
\cdot {\bf u}_t=0$.


\begin{thebibliography}{99} 

\bibitem{Heisenberg} W. Heisenberg,
Zeit. Phys. {\bf 33}, 879 (1925).

\bibitem{Schrodinger} E. Schr\"odinger, 
Annalen der Physik {\bf 79}, 361  (1926).

\bibitem{Groenwald} H.J. Groenwald, Physica {\bf 12}, 405 (1946).

\bibitem{van Hove} van Hove,
Mem. Acad. Roy. Belg. {\bf 26}, 61 (1951).

\bibitem{EPR}
A. Einstein, B. Podolsky, and N. Rosen, Phys. Rev. {\bf 47}, 777 (1935).

\bibitem{Bell}
J.S. Bell, Physics {\bf 1} (1964) 195;
{\it Speakable and unspeakable in
quantum mechanics}
(Cambridge University Press, 1987).

\bibitem{Aspect}
A. Aspect, P. Grangier, and G. Roger, Phys. Rev. Lett. {\bf 49}, 91 (1982).

\bibitem{Bohr} 
N. Bohr,
Dialectica  {\bf 1}, 312 (1948).

\bibitem{Ono} T. Ono,
Phys. Lett. A {\bf 230}, 253 (1997)./
the doctoral dissertation in University of Tokyo (1997).

\bibitem{Jung&Pauli}
C.G. Jung and W. Pauli, {\it The interpretation of Nature and the Psyche},
(Bollingen Foundation Inc., New York, 1955).

\bibitem{SbMII} T. Ono, 
Found. Phys., to be submitted.

\bibitem{SbMIII} T. Ono,
Found. Phys., to be submitted.

\bibitem{Wald} R.M. Wald,
{\it General Relativity},
(Univ. Chicago Press, USA, 1984).

\bibitem{Omori} H. Omori,
{\it Infinite-Dimensional Lie Groups},
(AMS, Providence, Rhode Island, 1997),
Trans. Math. Mono. Vol.158./
(Kinokuniya Co., Ltd. Japan, 1979),
in Japanese.

\bibitem{Kaluza}
Von TH. Kaluza, 
in: {\it Modern Klauza-Klein Theories}, ed. T. Appelquist,
A. Chodos, and P.G.O. Freund
(Addison-Wesley, USA/Canada, 1987).


\bibitem{Whitehead}   A.N. Whitehead, 
{\it Process and Reality}, 
edited by D.R. Griffin and D.W. Sherburne 
(The Free Press, New York and London, 1979).

\bibitem{Bogolubov}
N.N.
Bogolubov,
A.A. Logunov,
A.I. Oksak, and I.T. Todorov,
{\it General Principles of Quantum Field Theory},
(Kluwer, 1990).


\bibitem{K&N} S. Kobayashi and K. Nomizu,
{\it Foundations of Differential Geometry}
 (John Wiley, New York, 1963).

\bibitem{Lie} S. Lie, {\it  Theorie der
Transformationsgruppen} (Zweiter Abschnitt, Teubner, 1890).

\bibitem{Poincare} H. Poincar\'e,  
C.R. Acad. Sci. {\bf 132}, 369 (1901). 

\bibitem{M&R} J.E. Marsden and T.S. Ratiu,
{\it Introduction to Mechanics and Symmetry:
A Basic Exposition of Classical Mechanical Systems}
(Springer-Verlag, New York, 1994).

\bibitem{M&W2} J.E. Marsden and A. Weinstein,
Rep. on Math. Phys. {\bf 5}, 121  (1974).

\bibitem{G&S} V. Guillemin and S. Sternberg, 
Ann. Phys. {\bf 127}, 220  (1980)./
{\it Symplectic Techniques in Physics}
(Cambridge Univiversity Press, New York, 1991).

\bibitem{A&M} R. Abraham and J. Marsden, {\it Foundation of
Mechanics}
(Addison-Wesley, Reading,
MA, 1978), second edition, 

\bibitem{E&M} D. Ebin and J. Marsden, 
Ann.
Math. {\bf 90}, 102 (1970). 

\bibitem{Arnold} V.I. Arnold, 
 Ann. Inst. Fourier  
{\bf 16}, 319 (1966); {\it Mathematical Methods
of Classical Mechanics} (Springer, New York, 1978),
 Appendix 2. 

\bibitem{H&K} D. Holm and B. Kupershmidt,
Physica D {\bf 6},
347 (1983).

\bibitem{M&R&W} J. Marsden, T. Ratiu, and A.
Weinstein, Trans. Am. Math. Soc. {\bf 281}, 147  (1984).




\end{thebibliography}
\end{document}